\documentclass[useAMS,usenatbib]{mnras}
\usepackage{newtxtext,newtxmath}
\usepackage[T1]{fontenc}
\usepackage{ae,aecompl}
\usepackage{float}
\usepackage{graphicx}	
\usepackage{amsmath}	
\usepackage{hyperref}
\graphicspath{{./fig/}} 
\usepackage[normalem]{ulem}
\usepackage{xcolor} 
\usepackage{appendix}
\usepackage{chngcntr}

\newcommand{\ctworay}{\texttt{C$^2$RAY}~}
\newcommand{\secref}[1]{\S~\ref{#1}}

\defcitealias{Bianco2021segunet}{Paper I}

\title[SegU-Net v2]{Deep learning approach for identification of {\sc Hii} regions during reionization in 21-cm observations -- II. foreground contamination}
\author[M. Bianco et al.]{Michele Bianco,$^{1,2}$\thanks{Contact e-mail: michele.bianco@epfl.ch} Sambit. K. Giri,$^{3,4}$ David Prelogovi\'c$,^{5}$ Tianyue Chen,$^{1}$ Florent G. Mertens,$^{6}$\newauthor Emma Tolley,$^{1}$ Andrei Mesinger,$^{5}$ and Jean-Paul Kneib$^{1}$\\
$^{1}$ Laboratoire d’Astrophysique, Ecole Polytechnique Federale de Lausanne (EPFL), Observatoire de Sauverny, Versoix 1290, Switzerland\\
$^{2}$ Astronomy Centre, Department of Physics \& Astronomy, Pevensey III Building, University of Sussex, Falmer, Brighton, BN1 9QH, United Kingdom\\
$^{3}$ Nordita, KTH Royal Institute of Technology and Stockholm University, Hannes Alf\'vens v\"ag 12, SE-106 91 Stockholm, Sweden\\
$^{4}$ Institute for Computational Science, University of Zurich, Winterthurerstrasse 190, 8057 Zurich, Switzerland \\
$^{5}$ Scuola Normale Superiore, Piazza dei Cavalieri 7, 56126 Pisa, Italy\\
$^{6}$ LERMA, Observatoire de Paris, PSL Research University, CNRS, Sorbonne Universit\'e, F-75014 Paris, France}

\date{Accepted 2024 January 18. Received  2023 December 11; in original form 2023 April 6; Report Number NORDITA 2023-013}

\pubyear{2023}

\begin{document}
\label{firstpage}
\pagerange{\pageref{firstpage}--\pageref{lastpage}}
\maketitle
\begin{abstract}
The upcoming Square Kilometre Array Observatory (SKAO) will produce images of neutral hydrogen distribution during the epoch of reionization by observing the corresponding 21-cm signal. However, the 21-cm signal will be subject to instrumental limitations such as noise and galactic foreground contamination which pose a challenge for accurate detection. In this study, we present the \texttt{SegU-Net v2} framework, an enhanced version of our convolutional neural network, built to identify neutral and ionized regions in the 21-cm signal contaminated with foreground emission. We trained our neural network on 21-cm image data processed by a foreground removal method based on Principal Component Analysis achieving an average classification accuracy of 71 per cent between redshift $z=7$ to $11$. We tested \texttt{SegU-Net v2} against various foreground removal methods, including Gaussian Process Regression, Polynomial Fitting, and Foreground-Wedge Removal. Results show comparable performance, highlighting \texttt{SegU-Net v2}'s independence on these pre-processing methods. Statistical analysis shows that a perfect classification score with $AUC=95\%$ is possible for $8<z<10$. While the network prediction lacks the ability to correctly identify ionized regions at higher redshift and differentiate well the few remaining neutral regions at lower redshift due to low contrast between 21-cm signal, noise and foreground residual in images. Moreover, as the photon sources driving reionization are expected to be located inside ionised regions, we show that \texttt{SegU-Net v2} can be used to correctly identify and measure the volume of isolated bubbles with $V_{\rm ion}>(10\, {\rm cMpc})^3$ at $z>9$, for follow-up studies with infrared/optical telescopes to detect these sources.
\end{abstract}

\begin{keywords}
cosmology: dark ages, reionization, first stars, early Universe -- techniques: image processing, interferometric
\end{keywords}
\section{Introduction}
Radiation emitted by the first luminous sources drastically influenced the chemical composition and thermal history of the intergalactic medium (IGM), transitioning the Universe from an initial cold and neutral state to a final hot and ionized state \citep[e.g.][]{Furlanetto2006CosmologyUniverse, Ferrara2014, Choudhury2022}. These sources most likely formed at locations where dark matter structures collapsed into gravitational bound structures during redshift $z\gtrsim 10$  \citep{Abel2001TheUniverse, Bromm2009, Pawlik2011}. The newly launched \textit{James Webb Space Telescope (JWST)}\footnote{\url{http://jwst.nasa.gov}} is already providing preliminary results by detecting possible ionizing source candidates at these high redshifts \citep{Castellano2022, Naidu2022, Bakx2022}, which will help us understand the conditions for early galaxy formation \citep[e.g.][]{BoylanKolchin2022stress,hutsi2023did,Dayal2023wdm}.

Another way to probe the appearance of these first luminous sources is to observe the evolution of neutral hydrogen ({\sc Hi}) in the IGM. The ground state spin-flip transition of neutral hydrogen produces a signal with a wavelength of 21 cm in the rest frame, known as the \textit{21-cm signal}. The presence of this signal is directly correlated with the number density of neutral hydrogen present in the early Universe, and with the Universe expansion, the 21-cm signal wavelength redshifts into the radio frequency. As the first stars and galaxies formed and began emitting ultraviolet radiation, they started to ionize neutral gas in their surrounding. These primordial sources produce enough photons to escape their hosting environment and propagate into the IGM. As the hydrogen in the IGM becomes ionized, the intensity of the 21-cm signal decreases. Therefore, by observing the 21-cm signal from the early Universe with radio telescopes, we can study the reionization process and learn about the properties of the first luminous sources \citep[e.g.][]{Madau199721Redshift, Furlanetto2006CosmologyUniverse}. Several radio experiments, such as the Low-frequency Array\footnote{\url{https://www.astron.nl/telescopes/lofar}} \citep[LOFAR; e.g.][]{Mertens2020ImprovedLOFAR, Ghara2020ConstrainingObservations}, Murchison Wide-field Array\footnote{\url{https://www.mwatelescope.org}} \citep[MWA; e.g.][]{Trott2020DeepObservations, Ghara2021Constraining} and Hydrogen Epoch of Reionization Array\footnote{\url{https://reionization.org/}} \citep[HERA; e.g.][]{HERA2022UpperPhaseI, HERA2022UpperPhaseIcontraints}, have been trying to detect this signal during the epoch of reionization (EoR).

Currently, the low-frequency band component of the Square Kilometre Array\footnote{\url{https://skatelescope.org}} \citep[SKA-Low; e.g.][]{Mellema2013ReionizationArray}, which will observe the sky at a frequency range between $50$ and $350\,\rm MHz$, is under construction. SKA-Low will have a field of view covering $\sim$(10 $\rm deg)^2$ on the sky \citep{Koopmans2015TheArray}. This radio interferometer will be sensitive enough to capture the evolution of the IGM during EoR with images of the 21-cm signal from redshift $z=30$ to $5$. This sequence of 21-cm maps observed at different frequencies will be stuck together to constitute a three-dimensional set of data, known as the multi-frequency tomographic dataset \citep[e.g.][]{Mellema2015HISKA, Wyithe2015ImagingSKA, Giri2018BubbleTomography}. The 21-cm signal image data produced by the SKA-Low will contain imprints of the ionised regions (or bubbles) caused by the luminous sources \citep{Giri2018BubbleTomography, Giri2018OptimalObservations} and neutral regions (or islands) tracing the cosmic voids \citep{Giri2019NeutralTomography}. By detecting these bubbles, we can learn about the locations of the first luminous sources \citep{Zackrisson2020Bubble6}. We can also understand the nature and distribution of the photon sources driving the reionization process by studying the evolution of their sizes and morphology \citep[e.g.][]{Giri2018BubbleTomography, Giri2019NeutralTomography, Giri2020Betti, Kapahtia2019Morphology, Kapahtia2021, Gazagnes2021Inferring, elbers2022persistent}. However, detecting these ionised bubbles in radio telescope observations is not trivial due to several limitations of the telescope, such as the limited resolution and instrument noise. 

To detect these bubbles, previous authors have developed methods using visibilities data smoothed with appropriated filters to represent the sizes and shapes of the bubbles, then a likelihood for Bayesian approach estimates the parameters of the ionized regions filtered \citep[e.g.][]{Datta2007Detecting, Ghara2020BayesianBubbles}. Other authors employ the image data of radio telescopes. This approach can be intensity-based, where the method filters the image based on a threshold value or region-based, by agglomerate clustering correlated pixels into groups with common traits within the image \citep[e.g.][]{achanta2012SLICMethods, Mehra2016AMethods, Giri2018OptimalObservations}. This task is a well-known assignment in Artificial intelligence (AI) called segmentation. Therefore, another approach would be to consider a deep learning application. Recent work by \cite{GagnonHartman2021} demonstrated that a combination of foreground avoidance and machine learning techniques enable 21-cm segmentation and bubble detection for experiments that are not necessarily optimized for imaging. Moreover, recently, we presented our first work \cite[see][hereafter Paper I]{Bianco2021segunet}, where we introduced a deep learning approach to identify the distribution of {\sc Hi} regions in SKA 21-cm tomographic image using a U-shaped convolutional neural network (U-Net) \citep{Ronneberger2015}. We named our framework \texttt{SegU-Net} and we assessed how this network could process 21-cm images during the EoR contaminated by systematic noise simulated for SKA-Low and segment the images into ionized and neutral regions with an average of $87\%$ accuracy for redshift between 7 and 9. Moreover, we assessed that our network outperforms the \texttt{Super-Pixel} method \citep{Giri2018BubbleTomography}, considered the state-of-the-art algorithm for EoR segmentation, with, on average, $10$ to $20\%$ more accuracy. We also demonstrated that \texttt{SegU-Net} could be used to recover the bubble size distributions with a relative difference within the $5\%$ and other summary statistics with the same level of accuracy. Moreover, we provided our method with a per-pixel uncertainty map that provides a confidence interval for its prediction and the derived statistics. We have tested the response of our framework to different noise levels based on a shorter (250 h) and more extended (1500 h) observing time, corresponding to an under- and overestimation of the noise level, respectively. We demonstrated that \texttt{SegU-Net} tolerates noise up to $\sqrt{2}$ times larger than the one employed in the training process, obtaining the same level of accuracy. By studying the uncertainty map and the response to the noise level, we realised that machine learning models are sensitive to the dynamic range and the intrinsic resolution of the simulated images.

While our previous work demonstrated excellent performance in detecting {\sc Hi} regions from EoR images, it should be considered a proof-of-concept as we consider EoR images with only telescope systematic noise, and we did not include any foreground contamination. The biggest challenge for the SKA-Low observation, just like other radio telescopes, is to separate the 21-cm signal from the undesired extra-galactic and galactic foreground contamination, which outshine the cosmological signal by several orders of magnitude \citep{Jelic2008Foreground, Bowman2009}. The key goal of this work is to develop tools which remove these foregrounds while recovering the regions of {\sc Hi} during EoR from the 21-cm signal image data. 

In this work, we will further develop our deep learning-based method to determine the ionised bubbles in image data with the presence of realistic galactic and extra-galactic foregrounds expected from the SKA-Low. Therefore, here we present \texttt{SegU-Net v2}, which extends the previous work by including foreground emissions of galactic origin and a complete study of its dependency on the foreground mitigation pre-processing step that partially subtracts the foreground signal, thus reducing the dynamic range in the 21-cm images before starting the network training. In the last three decades, several foreground removal methods with different approaches have been developed. Some of the early attempts take advantage of the spectral smoothness of the galactic and extra-galactic contaminants to fit along the line of sight and remove the foreground in either real or $uv$ space \citep[e.g.:][]{Morales2006Stat, Morales2006Improv, Wang2006foreg, Gleser2008Decont, Liu2009Improv, Wang2013Exploring}. However, more recent approaches suggest a non-parametric subtraction \citep[e.g.][]{Harker2009Nonparametric, Gu2013Wavelet, Chapman2012Foreground, Chapman2013Scale, Bonaldi2015Foreground, Mertens2018Statistical21} as the frequency smoothness of the foreground spectrum can be corrupted by beam effect and incomplete $uv$ coverage \citep{Liu2009Will}. Therefore, we perform a complete study of different available approaches for foreground subtraction in the case of the SKA-Low 21-cm tomographic dataset applied to \texttt{SegU-Net v2}. We analyse the effect of the subtraction process on the predicted binary maps so that we can establish if a particular foreground removal method provides a concrete advantage for our task.

This paper is organised as follows. In \secref{chap:21cmsignal}, we present how we generate the simulated data sets used for this work, including details of our foreground model in \secref{sec:frg} and a description of the mock 21-cm observation in \secref{sec:mock_obs}. In \secref{chap:unetfor21cmimages}, we describe the design and the training of our neural network. In \secref{chap:results}, we discuss its application to our simulated SKA-Low data sets contaminated by the foreground signal, and we analyse summary statistics such as the mean ionization fraction, power spectra and topological quantities. In \secref{sec:other_preproc} we test our framework on a different foreground removal method. We discuss and summarize our conclusions in \secref{chap:discussion}. Throughout this work, we assume a flat $\Lambda$CDM cosmology with the following parameters: $\Omega_{\Lambda}=0.73$, $\Omega_{m}=0.27$, $\Omega_b=0.046$, $ H_0=70\,\rm{km\,s^{-1}{\rm Mpc}^{-1}}$, $\sigma_8=0.82$, $ n_s=0.96$. These values are based on the WMAP 5 years observation \citep{Komatsu2009} and consistent with \textit{Planck 2018} \citep{PlanckCollaboration2018} results.

\section{21-cm signal} \label{chap:21cmsignal}
This section illustrates the process we follow to create 21-cm mock observations of the EoR. Development of the network requires mock 21-cm observations of the EoR for network training, validation and testing, which will be described in \secref{chap:unetfor21cmimages}. 

\subsection{Simulating the Cosmological 21-cm Signal during EoR}\label{sec:dTb}
The intensity of the redshifted 21-cm signal emerging from a neutral cloud of hydrogen can be observed by a radio interferometric telescope as the difference against the CMB temperature $T_{\rm CMB}$, i.e. $\delta T_b \equiv T_b - T_{\rm CMB}$. For a given sky angular position $\pmb{\hat{n}}$ and redshift $z$, we can define it to be \citep[e.g.][]{Zaroubi2012EoR, Mellema2013ReionizationArray}
\begin{eqnarray}\label{eq:dTb}
   \delta T_\mathrm{b} (\pmb{r}, z) = T_0(z) \left(1-\frac{T_{\rm CMB}(z)}{T_{\rm S}(\pmb{r}, z)}\right)[1+\delta_b(\pmb{r}, z)]x_{\rm HI}(\pmb{r}, z), \\
   T_0 (z) \approx 27~\rm{mK} \left(\frac{\Omega_b}{0.044}\right)\left(\frac{h}{0.7}\right) \sqrt{\left(\frac{1+z}{10}\right)\left(\frac{0.27}{\Omega_m}\right)}\,.
\end{eqnarray}
where $ x_\mathrm{HI}$ is the neutral hydrogen fraction, $\delta_\mathrm{b}$ is the baryonic over-density, and $T_{\rm S}$ is the spin temperature. We assume that the IGM is heated well above the CMB temperature ($T_{\rm S}\gg T_{\rm CMB}$) at $z\lesssim 12$, which is consistent with theoretical predictions \citep[e.g.][]{Pritchard200721-cmReionization,Ross2017SimulatingDawn, Ross2019EvaluatingDawn,Ross2021CosmicDawn} \footnote{Note that the current 21-cm signal measurements have not completely ruled out the possibility of cold reionization \citep[see e.g.][]{Ghara2020ConstrainingObservations,Ghara2021Constraining, HERA2022UpperPhaseIcontraints}. The signal becomes very complicated if $T_{\rm S}\sim T_{\rm CMB}$ when reionization begins \citep{Ross2021CosmicDawn,Schneider2023CosmologicalReionization}. Therefore, we defer a detailed exploration to the future.}.
In this context, \autoref{eq:dTb} is always positive and can be approximated as $\delta T_b\propto (1+\delta_b)\,x_{\rm HI}$, while the presence of ionized regions is characterized by a lack of signal, $\delta T_b = 0\,\rm{mK}$. The radio interferometer cannot observe the absolute $\delta T_b$. Therefore, the ionised regions cannot be identified by finding pixels with zero signal in the 21-cm image data.
\begin{figure*}
	\includegraphics[width=0.9\textwidth]{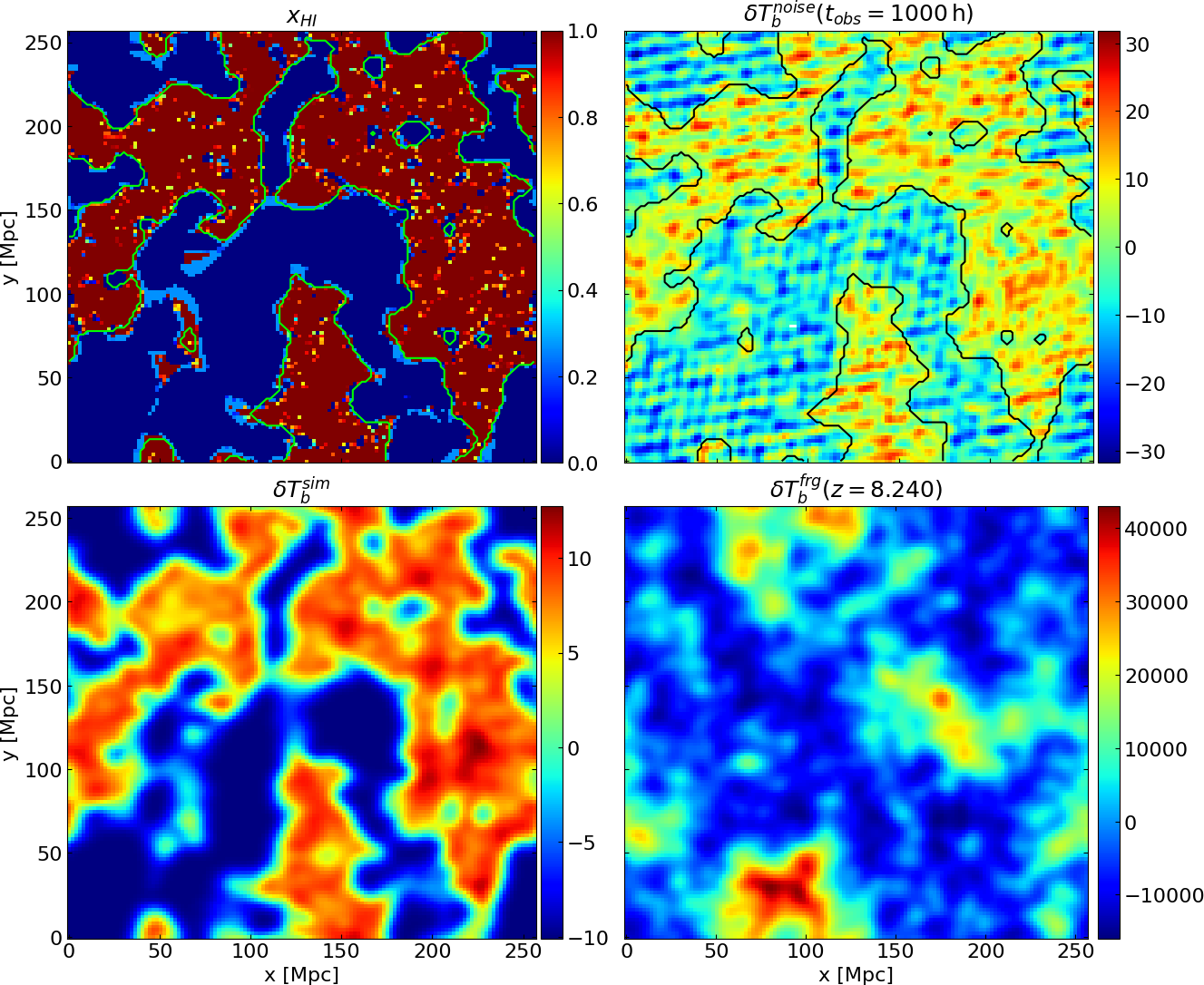}\vskip-2mm
	\caption{An example of a slice through the sky-plane used during the network training. \textit{Top Left}: the neutral hydrogen fraction at simulation resolution when the reionisation process is halfway complete. \textit{Bottom Left}: the simulated 21-cm signal after the interferometric smoothing with a maximum baseline of $B=2\,\rm km$ and matching frequency resolution. We then subtract the frequency mean signal to mimic the effect of the lack of a zero baseline. \textit{Top Right}: systematic noise added to the 21-cm signal for an observing time of $1000$ hours. A solid black line indicates the neutral field after the same interferometric smoothing scale. \textit{Bottom right}: the Galactic synchrotron emission added to the 21-cm signal with the systematic. We can notice how the dynamic range is a few orders of magnitude larger and completely outshines the 21-cm signal. For all the differential brightness images, the units are in $\rm{mK}$.}
	\label{fig:train_example}
\end{figure*}
To model the large-scale cosmological 21-cm signal expected during reionisation, we employ the \texttt{Python} wrapper of the \texttt{21cmFAST} semi-numerical code \citep{Mesinger2011, Murray202021cmFAST}. The code models the dark matter density evolution and gravitational collapse using the second-order Lagrangian perturbation theory (2LPT). From the generated large-scale density field, a region is considered collapsed when it exceeds a defined mass threshold, which can be related to a minimum virial temperature $T^{\rm min}_{\rm vir}$. The excursion set formalism is then employed to calculate ionised regions \citep{Furlanetto2004}. The code outputs a coeval cube at different redshifts that are then used for constructing 21-cm lightcones. We refer the readers to e.g. \citet{Giri2018BubbleTomography} for more general details on the construction of lightcone from coeval cube simulations. In this work, we simulate the signal in coeval cubes for a total of $\sim 20$ snapshot for redshift $z=\left[7, 11 \right]$ with a mesh grid of $128^3$ that is $256$ Mpc along each direction.

\subsection{Systematic Noise}\label{sec:noise}
We model the SKA-Low antenna receiver noise by a random Gaussian distribution with mean value zero and variance \citep{Ghara2017ImagingSKA, Giri2018OptimalObservations}
\begin{equation}\label{eq:syst_noise}
    \sigma_{\rm uv} = \frac{k_{\rm B}\, T_{\rm sys}}{A_{\rm eff}} \sqrt{\frac{2\,t_{\rm daily}}{\Delta\nu\,N_{\rm uv}\,t_{\rm obs}\,t_{\rm int}}}\,.
\end{equation}
Here $t_{\rm int}$ is the integration time, $t_{\rm daily}$ is the window of observation per day, $T_{\rm sys}$ is the system temperature, $A_{\rm eff}$ is the effective collecting area, $\Delta\nu$ is the bandwidth, $N_{\rm uv}$ is the number of measurements that are detected in each cell of the \textit{uv}-coverage grid. We assume an observation length of $t_{\rm obs}=1000\,\rm{h}$. We list the SKA-Low telescope parameters in \autoref{tab:telescope_param}. The \textit{uv}-coverage grid is simulated assuming the current plan for antennae distribution of SKA-Low\footnote{The SKA-Low design is given at \url{https://www.skao.int/sites/default/files/documents/d18-SKA-TEL-SKO-0000422_02_SKA1_LowConfigurationCoordinates-1.pdf}.}. In the top-right panel of \autoref{fig:train_example}, we show an example slice of the 21-cm signal and a noise realisation at $z=8.24$. As the map is degraded to a resolution corresponding to a maximum baseline of $B=2\,\rm km$, we can see the large-scale distribution of the neutral and ionised regions.

\begin{table}
	\centering
	\caption{The telescope parameters used in this work. For the frequency channel width, we indicate the quantity at $z=7$ and $11$.}
	\label{tab:telescope_param}
	\begin{tabular}{lcc} 
		\hline
		Parameters & Values \\
		\hline
        System temperature & $T_{\rm sys}$ & $60 (\frac{\nu}{300\mathrm{MHz}})^{-2.55}$ K  \\
		Effective collecting area & $A_{\rm eff}$ & 962 $\mathrm{m}^2$  \\
        Declination & $\theta_c$ & -30$^\circ$ \\
        Frequency channel width & $\Delta\nu$ & $118 - 96\,\rm{kHz}$\\
        Observation hour per day & $t_{\rm daily}$ & 6 hours \\
        Signal integration time & $t_{\rm int}$ & 10 seconds \\
		\hline
	\end{tabular}
\end{table}

\subsection{Foreground Contamination}\label{sec:frg}
Between $250$ and $30\,\rm{MHz}$, the dominant emission comes from the Galactic synchrotron radiation. This emission alone is expected to contribute to the majority of the total foreground contamination of the comic 21-cm signal \citep{DiMatteo2002, DiMatteo2004, Santos2005Multifreq, Datta2007Detecting, Jelic2008Foreground, Kerrigan2018}. Other contributors can include emissions from unresolved extra-galactic point sources, Galactic free–free emissions, supernova remnants and extra-galactic radio clusters, which, for simplicity, have been neglected in this study. We based our Galactic synchrotron emission model on the \cite{Choudhuri2014foreground} study. We express the foreground radiation with a Gaussian random field with an angular power spectrum as
\begin{equation}\label{eq:foreground_powerspect}
    \mathcal{C}^{\rm syn}_l(\nu) = A_{150}\, \left(\frac{1000}{l}\right)^{\overline{\beta}} \left(\frac{\nu}{\nu_{\star}}\right)^{-2\overline{\alpha}_{\rm syn} -2 \Delta\overline{\alpha}_{\rm syn}\,\rm{log} \left(\frac{\nu}{\nu_{\star}}\right)}\,.
\end{equation}
Here, the parameter for the Galactic synchrotron emission is the power spectra amplitude $A_{150}=512\,\rm{mK}^2$ at the reference frequency $\nu_{\star}=150\,\rm{MHz}$, the angular scaling $\overline{\beta}=2.34$, the spectra index $\overline{\alpha}_{\rm syn}=2.8$ and the running spectral index $\Delta\overline{\alpha}_{syn}=0.1$. These quantities are taken from \cite{Platania1998}, and \cite{Wang2006foreg}. We then generate the foreground temperature fluctuations map following the relation 
\begin{equation}\label{eq:frg_map}
    \delta T_b^{\rm frg}(U,\,\nu) = \sqrt{\frac{\Omega_{\rm SKA}\,\mathcal{C}^{\rm syn}_l(\nu)}{2}} \left[x_l(U) + i\cdot y_l(U) \right]\,.
\end{equation}
$\Omega_{\rm SKA}$ is the total SKA-Low solid angle and $U = l /2\pi$. The two quantities $x_l$ and $y_l$ are independent random Gaussian variables with mean zero and variance of one, $\mathcal{N}\sim(0, 1)$. By performing two-dimensional inverse fast-Fourier transform of \autoref{eq:frg_map}, we get the spatial distribution of the foreground contamination $\delta T_b^{\rm frg}(\pmb{\hat{n}},\,z)$. With each lightcone simulation, we fix the random variables seed for the lowest redshift, $z=7$, and compute \autoref{eq:foreground_powerspect} for the corresponding frequency of the image.

\subsection{Mock 21-cm Observation} \label{sec:mock_obs}
From the simulated coeval cubes described in \S\ref{sec:dTb}, we create 3D lightcones with differential brightness $\delta T^{\rm sim}_b(\pmb{\hat{n}},z)\equiv\delta T^{\rm sim}_b(x,y,z)$ at $x$, $y$ coordinates for a total box size of $256\,\rm cMpc$ and spatial resolution of $\Delta x = 2\,\rm cMpc$, both in comoving units, corresponding to an angular mesh-size of $128^2$. This scale corresponds to an angular resolution of $\Delta \theta = 0.77 \,\rm arcmin$ at redshift $z=7$. The redshift coordinate is divided into 552 bins at equal comoving distance $\Delta x$ from $z=11$ to 7, corresponding of frequencies from $\nu_{\rm obs} = 118\,\rm MHz$ to $178\,\rm MHz$ and a frequency resolution of approximately $\Delta \nu \simeq 0.11\,\rm MHz$.

We select one tomographic simulation from the prediction dataset as our \textit{fiducal} simulation. In \autoref{fig:train_example}, left column, we show a slice of this \textit{fiducial} lightcone at redshift $z=8.24$, corresponding to $\nu_{\rm obs}=152.90\,\rm MHz$. At this stage, the simulated lightcone is 50\% ionised. The top panel show the neutral fraction $x_{\rm HI}$, with blue and red regions being the neutral and ionised regions, respectively. At the same time, the green colour indicated regions of transitions with $x_{\rm} \simeq 0.5$. The differential brightness is calculated with \autoref{eq:dTb} with the approximation discussed in \S\ref{sec:dTb}. 

From radio interferometry telescope, we can obtain images by gridding the uv-plane and inverse Fourier transform the gridded visibility \citep{Smirnov2011rime, Offringa2014wsclean}. Image weighting can be applied to the visibilities before the gridding, and in the case of large-scales 21-cm EoR experiment with SKA-Low, the so-called natural weighting is preferable as the more redundant, short baselines ensures the highest signal-to-noise ratio in the image at the expense of a limited image resolution and large side lobes effect \citep{Briggs1995phd}. In our case, we do not simulate the 21-cm signal from the visibility space but instead work on images already in the real space. Therefore, to mimic the effect of the limited resolution due to the visibility weighting, in the angular direction, we apply a Gaussian kernel, $G(\pmb{\hat{n}},z)$, with Full-Width at Half Maximum (FWHM) of 
$21~\mathrm{cm}(1+z)/B$, where $B=2\,\rm km$ that corresponds to the maximum baseline of SKA-Low. According to the planned SKA-Low design\footnote{The construction document can be found at \url{https://www.skao.int/en/resources/402/key-documents}.}, it will be densely filled within this 2 km providing enough sensitivity to construct images. 
The bottom panel in \autoref{fig:train_example} shows the differential brightness after smoothing the field with $G(\pmb{\hat{n}},z)$. For reference, this interferometric smoothing corresponds to an angular resolution of $\sim2.9\,\rm arcmins$ at $z\approx 7$ and $\sim 4.3\,\rm arcmins$ at $z\approx 11$. In the frequency direction, we apply a top-hat bandwidth filter with the same width as the FWHM in the angular direction. We implement the method explained in \S\ref{sec:noise} and the parameters listed in \autoref{tab:telescope_param} to simulate the effect of the systematic noise, $\delta T^{\rm noise}_b(\pmb{\hat{n}}, z)$. We create a random field with the same mesh size as the lightcone and add the simulated differential brightness. We then apply the same interferometric smoothing mentioned above, and the result is shown in \autoref{fig:train_example}, top right panel. As a reference for the reader, this was the network input in our previous work \citepalias{Bianco2021segunet}.

In this paper, we want to extend our previous effort as we want to recover the neutral binary map in the presence of contamination due to the synchrotron Galactic foreground, $\delta T^{\rm frg}_b(\pmb{\hat{n}}, z)$. The result of the model described in \S\ref{sec:frg} is shown in \autoref{fig:train_example}, bottom right panel. As we can see, the dynamic range of the observed changes drastically. Our previous work showed that our method is sensitive to the SNR level between the noise and the 21-cm signal. Therefore, we need to introduce an additional pre-processing step in our framework to mitigate foreground contamination and decrease the dynamic range of the contaminated images before providing them for network training. We will discuss this method in more detail in \S\ref{sec:frg_mitigation}.

We can describe our mock observation pipeline by combining the components and operations described here above as \citep[e.g.][]{Liu2019Data}
\begin{equation}
	\delta T_{\rm obs}(\pmb{\hat{n}}, z) = \delta T^{\rm sim}_b(\pmb{\hat{n}}, z) + \delta T^{\rm frg}_b(\pmb{\hat{n}}, z) + \delta T^{\rm noise}_b(z)\,.
\end{equation}
For each realization of the lightcone $\delta T_{\rm obs}(\pmb{\hat{n}}, z)$, illustrated with \autoref{fig:train_example}, we calculate the mean along the frequency channels,
\begin{equation}
    \delta \overline{T}_{\rm obs}(z) = \frac{1}{N_x N_y} \sum_{i=1}^{N_x} \sum_{j=1}^{N_y} \delta T_{\rm obs}(x_i,y_j,z) \, ,
\end{equation}
where $N_x$ and $N_y$ are the dimension in the angular-direction of the $128^2$ mesh. We subtract this quantity from $\delta T_{\rm obs}$ to account for the effect of the null baseline in interferometry telescopes. For this reason, the colour bar in the figure shows a negative value. W convolve the subtracted term with the Gaussian kernel $G$ mentioned above
\begin{equation}\label{eq:dTobs}
    \delta \widetilde{T}_{\rm obs}(\pmb{\hat{n}},z) = \int_{\Omega_{\rm SKA}} \left[\delta T_{\rm obs}(\pmb{\hat{n}}',z) -  \delta \overline{T}_{\rm obs}(z) \right]\cdot G(\pmb{\hat{n}}-\pmb{\hat{n}}', z)\,d\pmb{\hat{n}}' \,.
\end{equation}
This result constitutes a realistic mock observation of the SKA-Low interferometric telescope, including systematic noise, Galactic foreground contamination, and telescope limited resolution effect. We employ this pipeline to create the training, validation and \textit{random testing set}. In \S\ref{sec:frg_mitigation}, we explain how we pre-process this type of data before inputting it into our neural network.

Finally, we create an additional field that serves as the target of the network training. We apply the interferometric smoothing explained above to the simulated neutral fraction field $x_{\rm HI}$ (top left panel \autoref{fig:train_example}). We then choose a threshold of $x_{\rm th} = 0.5$ to discern the ionised and neutral regions. The result is a binary lightcone, $x^B_{\rm HI}(\pmb{\hat{n}},z)$, where neutral and ionized regions are classified by $1$ and $0$, respectively. For a visual comparison, we over-plot the contour of this binary field as a black line in \autoref{fig:train_example} top right panel.

\begin{figure}
	\includegraphics[width=\columnwidth]{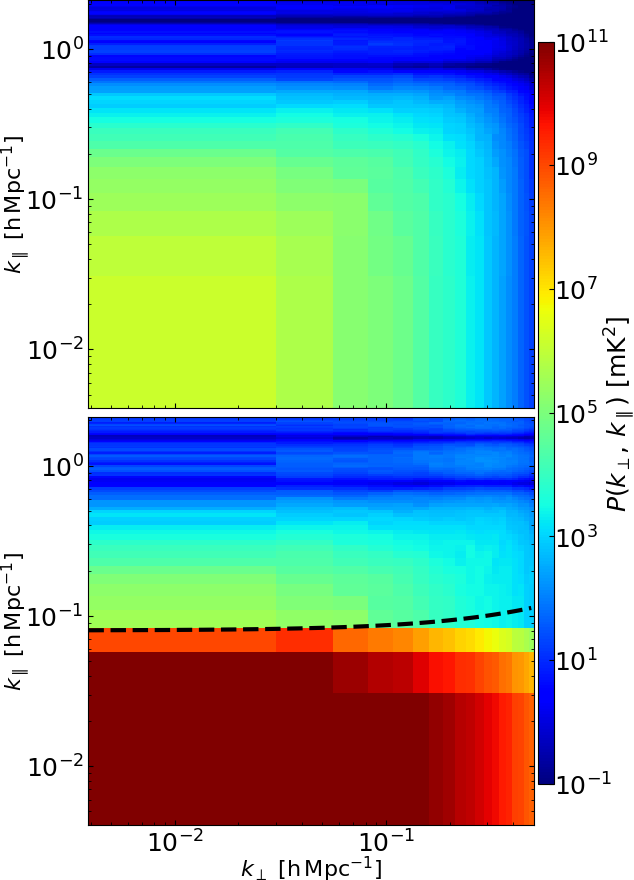}\vskip-2mm
	\caption{Cylindrical power spectra for a lightcone sub-volume centered at redshift $z_c=8.24$ and frequency depth of $\rm\pm 10\, MHz$. \textit{Top Panel}: 2D Power spectra from the simulated 21-cm signal only. \textit{Bottom Panel}: Same quantity but with the galactic foreground contribution. The black dashed line indicates the wedge slope with $\theta=2.25^{\circ}$ and $b=8\times10^{-2}\,\rm h\,Mpc^{-1}$.}
	\label{fig:ps_example}
\end{figure}

\begin{figure*}
	\includegraphics[width=\textwidth]{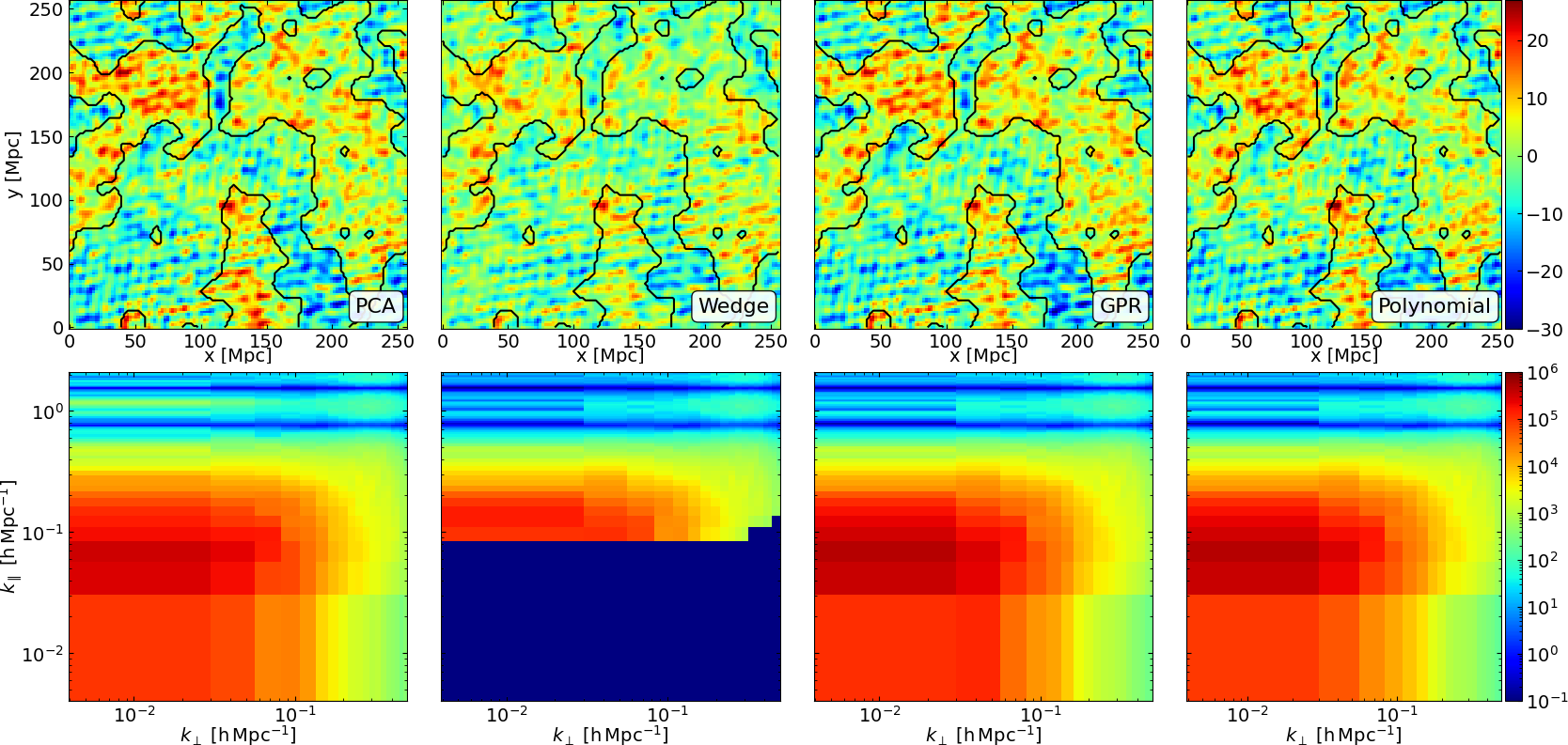}\vskip-2mm
	\caption{Comparison between different foreground mitigation methods. From left to right, we have PCA, wedge removal, GPR and polynomial fitting. First row, a visual example at redshift $z=8.24$ of the residual image after the corresponding method. Second row, the cylindrical power spectrum for a lightcone sub-volume centred at $z_c=8.24$ and frequency depth $\pm10\,\rm MHz$.}
	\label{fig:example_comparison}
\end{figure*}

\section{Foreground Mitigation}\label{sec:frg_mitigation}
As we outlined in \S\ref{sec:mock_obs}, foreground contamination poses a huge problem in detecting the 21-cm signal, as this signal is several orders of magnitude fainter in comparison. In \autoref{fig:ps_example}, we illustrate the effect of the foreground contamination on the 2D cylindrical power spectrum for a lightcone sub-volume centred at redshift $z_c=8.24$ and frequency width of $\Delta\nu \pm 10\,\rm MHz$. This quantity of the 21-cm signal (top panel) is compared with the same signal contaminated by the Galactic foreground signal (bottom panel). We observe that the contamination is visible at $k_{\parallel}\leq10^{-1}\,\rm Mpc/h$ with signal intensity of $\geq 10^9\,\rm mK^2$. The black dashed line in the figure indicates the foreground wedge. We will discuss this line later in \S\ref{sec:wedge}. To reduce the dynamic range of the foreground contaminated images to a level that is manageable for the neural network, we include a pre-processing step on the observed data, $\delta \widetilde{T}_{\rm obs}(\pmb{\hat{n}}, z)$. Hereafter, we refer to the resulting images of this pre-process as \textit{residual lightcone} or \textit{images}, $\delta T_{\rm res}(\pmb{\hat{n}}, z)$.

In foreground mitigation, we can consider two methods: foreground subtraction or avoidance \citep{Chapman2019ForegroundsMitigation}. Here, we consider three of the former cases, namely PCA, GPR and Polynomial fitting, and one of the latter techniques, Wedge removal. In this section, we briefly describe four different pre-processing methods that we test and we provide the residual image in \autoref{fig:example_comparison} for each method. The top panels show the residual image of the example illustrated in \autoref{fig:train_example}, while black contours indicate the ground truth. The bottom panel shows the 2D cylindrical power spectrum for the fiducial lightcone sub-volume centred at $z_c=8.24$ and frequency depth of $\pm 10\,\rm MHz$.

\subsection{Principal Component Analysis}\label{sec:pca}
Principal Component Analysis (PCA) is a commonly used method to remove foregrounds in 21-cm experiments~\citep[e.g.][]{Alonso2015BlindForeground,Cunnington2023HIintensity,Chen2023Detect}. The method exploits the fact that foregrounds have large amplitude and smooth frequency coherence. PCA simultaneously identifies the largest foreground components and an optimal set of basis functions that describe the frequency structure of the foregrounds. As the foregrounds are highly correlated in frequency, the frequency-frequency co-variance matrix of the foregrounds will have a particular eigensystem where most of the information can be sufficiently described by a small set of very large eigenvalues, the other ones being negligibly small. Thus, we can attempt to subtract the foregrounds by eliminating the components corresponding to the eigenvectors of the frequency co-variance matrix with the largest associated eigenvalues. In practice, we remove $4$ components, which captured most of the variance of the foreground modes. PCA is a relatively fast and computationally efficient method that requires no prior assumptions about the foregrounds or the 21-cm signal. However, PCA is not well-suited to handle non-linear relationships between the foregrounds and the 21-cm signal, and it can struggle to remove residual foregrounds not well-described by the largest components.

In \autoref{fig:example_comparison}, left column, we show the residual image at $z_c=8.24$, on top. After removing the first four components with PCA decomposition on the $20\,\rm MHz$ sub-volume of the fiducial lightcone, we obtain this image. On the bottom panel, we show the corresponding 2D power spectra.

\subsection{Wedge Remove}\label{sec:wedge}
We consider another pre-process that focuses on discarding the Fourier modes dominated by foreground contamination. This method assumes that the contaminated modes are contained in specific regions in the $k_{\perp}-k_{\parallel}$ space, named the \textit{foreground wedge}. These contaminated $k$-modes can be defined by \citep[e.g.][]{Liu2014EpochReduction, Murray2018}
\begin{equation}\label{eq:wedge}
    k_{\parallel} \leq |{\bf k}_{\perp}|\,\frac{H(z)}{1+z} \int^z_0\frac{dz'}{H(z')}\cdot \sin{\theta} + b \, ,
\end{equation}
where $H(z)$ is the Hubble parameter and ${\bf k}_{\perp}$ is the Fourier component perpendicular to the line of sight. $\theta$ is the angular size of the field of view, which can be interpreted as the horizon limit angle. $b$ is the bias that accounts for the presence of an intrinsic foreground limit at low $k_{\parallel}$-values. Pessimistic and arguably more realistic assumptions consider the horizon limit to $\theta=90^{\circ}$ justified by antenna side-lobes effect \citep{Pober2014What, Dillon2014Overcoming}. In our case, we select $\theta=2.25^{\circ}$, corresponding to the field of view (FoV), at redshift $z=7$ and comoving size of $256\,\rm cMpc$, of our dataset. We then select $b=8\times10^{-2}\,\rm h\,Mpc^{-1}$ based on the 2D cylindrical power spectrum shown in the right panel of \autoref{fig:ps_example}. The dashed black line indicates \autoref{eq:wedge} for the $\theta$ and $b$ mentioned before.

In this work, we employ a simplified version of the code developed by \cite{Prelogovic2021ML}. Here we give a brief description, referring the reader to the original paper for more details. First, we perform a 2D Fourier transform in the angular direction of a lightcone sub-volume, \autoref{eq:dTobs}, centred at redshift $z_c$ and with a given frequency depth, $\pm\Delta\nu$. Subsequently, an iterating procedure along the line-of-sight axis calculates \autoref{eq:wedge} and sets the $k$-modes that obey the condition to zero. A Blackmann-Harris taper function of the same angular and redshift size is multiplied by the lightcone to avoid artificial ringing in the Fourier space. However, this taper has the limitation that at low $k_{\parallel}$, it reduces the Fourier-space side lobes, while the opposite effect occurs at high $k_{\parallel}$. Finally, we do an inverse Fourier transform to regain the real-space lightcone sub-volume.

An example of data with the foreground contamination removed by this algorithm can be seen in the second column of \autoref{fig:example_comparison}. From the residual image (top panel), we see a large portion of the foreground residual is still present. The bottom panel shows the 2D cylindrical power. The dark blue colour indicates the $k_{\perp}-k_{\parallel}$ modes where the wedge removes method is applied.

\subsection{Gaussian Process Regression}\label{sec:gpr}
The Gaussian regression processes (GPR) method was developed in \cite{Mertens2018Statistical21} to separate foregrounds from 21-cm signal by modelling the two components as a stochastic process and separating them using a Bayesian approach. The method involves constructing a prior statistical model of the foregrounds and the 21-cm signal and then using the model to estimate the posterior distribution of the 21-cm signal given the observed data. This is done by assuming that the foregrounds and 21-cm signals are realizations of Gaussian processes, fully defined by their covariance. The selection of the prior covariance model in GPR is made under a Bayesian framework by maximizing the marginal likelihood. The Mat\'ern class of covariance functions is commonly used as prior covariance for the different data components. Following \citet{Mertens2018Statistical21}, a Radial Basis Function (RBF) kernel is used as the prior covariance model for the foreground component, while an Exponential kernel is used for the 21-cm signal. This method can effectively remove foreground contamination from the 21-cm signal and has the advantage of being able to incorporate prior knowledge about the signal and foregrounds. However, it requires accurate modelling of the foregrounds and assumptions about the statistical properties of the signal and foregrounds.

In \autoref{fig:example_comparison}, third column, we show the result obtained by the GPR presented here above. Similar to PCA, see \S\ref{sec:pca}, GPR removes a good portion of the foreground contamination providing a better contrast between the 21-cm emitting regions and the ionized one. For instance, the regions around $(x,y)=(225,\,100) \rm Mpc$ and $(x,y)=(150,\,200) \rm Mpc$. From the 2D power spectra at $k_{\parallel} > 3 \times 10^{-2}$ we see more signal when compared to PCA pre-process data.

\subsection{Polynomial fitting}\label{sec:poly}
We can also use Polynomial fitting to remove foreground contamination from the 21-cm signal \citep{Wang2006foreg, Alonso2015BlindForeground}. The method involves modelling the foregrounds as a smooth polynomial function in log space and fitting this function to the observed data, $\delta \widetilde{T}_{\rm obs}$.
\begin{equation}
    \log\left(T(\pmb{\hat{n}}, z)\right) = \sum^{N_{fg}}_{k=1} \alpha_k(\pmb{\hat{n}})\,\left[\log\left({\frac{\nu_0}{1+z}}\right) \right]^{k-1} \,.
\end{equation}
Here, $\nu_0$ is the 21-cm frequency and $N_{\mathrm{fg}}$ indicates the polynomial degree. In our study, we consider a fourth-degree polynomial. The resulting fit is then subtracted from the data to remove the foreground contamination $\delta T_{\rm res} = \delta \widetilde{T}_{\rm obs} - T(\pmb{\hat{n}}, z)$. 

This approach has the advantage of being simple and computationally efficient but may not be as effective at removing foregrounds as other, more sophisticated methods. One limitation of the polynomial fitting is that it assumes the foregrounds can be well-described by a smooth polynomial, which may not always be the case \citep[e.g.][]{Thyagarajan2015ForegroundsWide}. Additionally, if the polynomial fit is not in high enough order, it may leave some foregrounds in the data, while an overly high-order polynomial may also remove the signal. The polynomial fitting has been combined with other foreground removal methods in some studies to improve the overall performance of the foreground removal process.

In \autoref{fig:example_comparison}, fourth column, we show the result obtained by the Polynomial fitting. In both cases, from the residual image and the 2D power spectra, visually, we see similar results to GPR, see \S\ref{sec:gpr}, with a more considerable difference between the positive (neutral) and negative (ionized) regions in the residual image, although presenting the same level of residual foreground located at $(x,y)\sim(80,\,125)\,\rm Mpc$ as in the other methods.

\section{U-Net for 21-cm image segmentation}\label{chap:unetfor21cmimages}
The network architecture of \texttt{SegU-Net v2} is the same as in \citetalias{Bianco2021segunet}. The only implementation consists of a simplistic hyper-parameter optimization analysis on seven network hyper-parameters. In \S\ref{app:hyperpar}, we give a brief overview of the hyper-parameter space exploration method we employed and in \autoref{tab:hyperparams}, we list the six best-performing setups we found. Moreover, in \S\ref{app:blackbox},  we present a first attempt to open the \textit{black box} and performed a Gradient-weighted Class Activation Mapping (Grad-CAM) \citep{Selvaraju2019gradcam} importance analysis to highlight the features in the input image that the network employs to identify and predict the neutral regions from residual images. In \autoref{fig:inside}, we give a visual representation of the Grad-CAM importance analysis we performed.

\subsection{Network Architecture}
Here, we give a brief description of our network architecture. We refer the reader to our previous work for more details. \texttt{SegU-Net} is a U-shaped deep convolutional neural network composed of a contracting (encoder) and an expanding path (decoder). The former has two convolutional blocks, followed by the 2D averaging pooling operation of size $2^2$ and a dropout layer with a 5 per cent rate, \texttt{Encoder-Level=2*ConvBlock+AvrgPool+Drop}. A convolutional block consists of a 2D convolutional layer with kernel size $7^2$, followed by batch normalization and Rectified Linear Unit (\texttt{ReLU}) activation function, \texttt{ConvBlock=Conv2D+BN+ReLU}. The latter path consists of transposed 2D convolution followed by the concatenation with the corresponding output of the convolutional encoder block, dropout layer and two convolutional blocks, \texttt{Decoder-Level=TConv2D+CC+Drop+2*ConvBlock}. This structure is repeated four times for both the encoder and decoder. At each level, the pooling operation halves the angular dimension of the input and doubles the number of channels. The network takes as input a redshift slice from the residual lightcone, $\delta T_{\rm res}$, and outputs the corresponding 2D binary image, $x^B_{\rm HI}$.

\subsection{Dataset}
We generated a large set of realisations of the SKA multi-frequency tomographic dataset by changing the initial conditions and the following three astrophysical parameters. We sample the high-redshift galaxy efficiency $\zeta$ and the mean-free path of ionising photons $R_{\rm mfp}$ with a normal distribution with mean and variance $\mathcal{N}(82,\,18)$ and $\mathcal{N}(17.5\,\rm{Mpc},\,4.5\,\rm{Mpc})$, respectively. At the same time, the minimum virial temperature for star-forming halos $T^{\rm min}_{\rm vir}$ is sampled in logarithmic space with distribution $\mathcal{N}(4.7,\,0.2)$. We chose this sampling of parameters because we want the global volume-averaged neutral fraction $\overline{x}_{\rm HI}$ of all data to be at least greater than $90\%$ at redshift $z=11$ and less than $10\%$ at redshift $7$. Moreover, with this parameter sampling, we can postulate the spin-saturation assumption, $T_S \gg T_{\rm CMB}$, which assures that the differential brightness is strictly positive and that neutral hydrogen is correlated with a positive signal in each image.

In this work, we updated the dataset from \citetalias{Bianco2021segunet} for a total of 10,000 samples for the network training and 1,500 for validation. Once the network is trained, we will test its accuracy and generalisation ability on an additional 300 mock observations during the prediction step. We will refer to this dataset as the \textit{random testing set}. The training dataset is employed during the forward- and back-propagation \citep{Rumelhart1985}, while the validation dataset is used to validate the accuracy of network results during training. We want to clarify that we trained \texttt{SegU-Net v2} on $\delta T_{\rm res}$ data pre-processed only with the PCA eigen-decomposition on the full redshift range, $z=7$ to $11$, which is explained in \S\ref{sec:pca}. The testing dataset is an independent set of simulations on which we will validate the final results of the trained network. 
\begin{figure*}
	\includegraphics[width=1.05\textwidth]{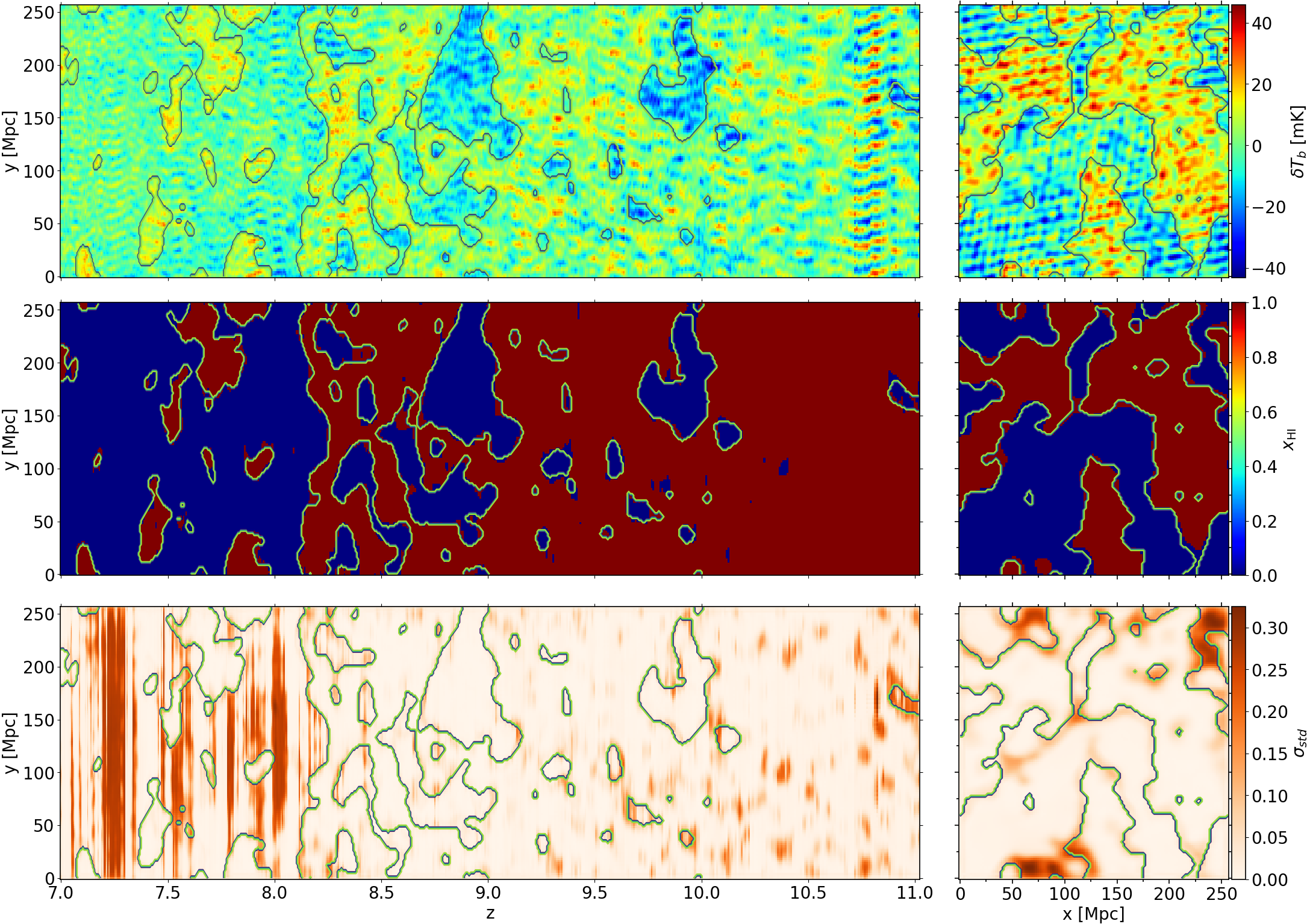}\vskip-2mm
	\caption{Visualisation of the different fields for our fiducial lightcone. \textit{Top Left}: for a given position on the x-direction, the redshift evolution of the residual lightcone after the PCA pre-processing step. \textit{Top Right}: residual image at redshift $z=8.24$ ($\overline{x}_{\rm HI}=0.5$). Same image as in \autoref{fig:train_example}. \textit{Middle Left}: redshift evolution of the predicted neutral (red) and ionised (blue) lightcones. \textit{Middle Right}: predicted map at the corresponding redshift. \textit{Bottom Left}: the corresponding per-pixel error lightcone, orange colour indicates the intensity of the uncertainty. \textit{Bottom Right}: the corresponding per-pixel error map. For all panels, we over-plot contours that represent the ground truth.}
	\label{fig:vis_eval}
\end{figure*}

\subsection{Metrics}
We consider a true positive detection ($TP$) to be the number of pixels correctly identified as neutral, while a true negative ($TN$) is the opposite. False positives ($FP$) and false negatives ($FN$) represent the number of pixels wrongly classified as neutral or ionised. Therefore, we can define the Matthews correlation coefficient (MCC) for quantifying the accuracy of our network predictions as 
\begin{equation}
    r_{\phi} = \frac{ TP \cdot TN - FP \cdot FN }{ \sqrt{ (TP+FP) (TP+FN) (TN+FP) (TN+FN) } }\,.
\end{equation}
This metric can have values between $-1\leq r_{\phi}\leq1$, quantifying the quality of binary field (two-class) classifications. A negative value indicates anti-correlation, zero represents a completely random classification, and positive values indicate a positive correlation. For a direct comparison with previous studies on segmentation of 21-cm image data \citep[e.g.][]{GagnonHartman2021}, we define three additional statistical metrics as follows
\begin{equation}
    {\rm Accuracy} = \frac{TP + TN}{TP+ FP + FN + TN}\,.
\end{equation}
Here, this metric indicates how well a model is able to predict the target variable correctly.
\begin{equation}
        {\rm Precision} = \frac{TP}{TP+ FP}\,.
\end{equation}
This second metric refers to the level of consistency or repeatability of a predicted value. While accuracy and precision are important metrics in evaluating the performance of a network, they may not be sufficient in certain scenarios. For instance, in our binary classification problem, there can be scenarios when neutral regions can be much rarer than ionised regions and vice versa. In this case, accuracy can be misleading as the model may achieve high accuracy by simply predicting the majority class for all instances. Precision and recall are more informative metrics in such cases as they consider the class imbalance.
\begin{equation}
    {\rm IoU} = \frac{TP}{TP+ FP + FN}\,.
\end{equation}
However, here, we include the third additional metric, the Intersection over Union (IoU), that quantifies how well the predicted neutral region of interest overlaps with the true one. We will use these metrics later in \S\ref{sec:other_preproc}. 

In our case, we are targeting binary maps that indicate the location in the sky at a given redshift as either neutral or ionized. Therefore, an easy way for the reader to interpret the results is in the number of pixels guessed correctly or wrongly. For this reason, we introduce the false positive rate (FPR), also referred to as non-specificity, and the true positive rate (TPR), also known as sensitivity.
\begin{equation}\label{eq:fpr_tpr}
    TPR = \frac{TP}{TP+FN}\hskip3mm , \hskip3mm FPR = \frac{FP}{FP+TN} .
\end{equation}
The former quantity gives the percentage of neutral pixels (positive case in our context) correctly identified as neutral. A value of $TPR=1$ will indicate that the network identified all the neutral pixels correctly. Otherwise, $1-TPR$ indicates the percentage of pixels falsely classified as ionized. Similarly, the $FPR$ gives the percentage of pixels falsely detected as neutral.

\subsection{Per-Pixel Error Estimation}
The error calculation uses the same method as in \citetalias{Bianco2021segunet}. In the prediction step, we employ temporal time augmentation (TTA) operations \citep{Perez2017, Wang2020} on the network input data to create several copies of the same realisation but that we modify by rotating and vertical/horizontal flip operation. In this work, we fix the axis of symmetry and rotation to the frequency direction. Thus, the number of manipulations was reduced to a sample of 16 copies. This number corresponds to the maximum independent operations we can apply to an image. \texttt{SegU-Net v2} then gives a prediction for each modified copy that is then rotated or flipped back to obtain a different prediction of the same input image. We calculated the standard deviation, $\sigma_{std}$, on the 16 copies and obtained a per-pixel uncertainty map as shown in \autoref{fig:vis_eval}, bottom panel. The method is simple but efficient, showing how difficult it was for the network to give the predicted binary field for each pixel in the image.

\section{Results}\label{chap:results}
This section discusses the result obtained with \texttt{SegU-Net v2} acting on data pre-processed with the PCA foreground removal method as explained in \S\ref{sec:pca}. Here, we evaluate the result on the predicted binary maps and the network performance on the different methods (illustrated in \S\ref{sec:frg_mitigation}) in \S\ref{sec:evaluation} and \S\ref{sec:other_preproc} respectively. Finally, in \S\ref{sec:ionbubble}, we demonstrate a possible astrophysical application of \texttt{SegU-Net v2}.

\subsection{Identifying {\sc Hii} Regions with \texttt{SegU-Net v2}}\label{sec:evaluation}
In \autoref{fig:vis_eval}, we visually evaluate one realisation of the network predicted neutral (red) and ionized (blue) regions. We refer to this simulated lightcone as the \textit{fiducial} simulation. In the right column, we show a slice at redshift $z=8.24$ ($\nu_{\rm obs}=152.90\,\rm{MHz}$), corresponding when the global volume average neutral fraction is $\overline{x}_{\rm HI}=0.5$. From top to bottom, we show the residual image after the PCA pre-processing employed as the input of the neural network, the binary map predicted with \texttt{SegU-Net v2} from the PCA pre-processed data and the derived per-pixel uncertainty, respectively. In the left column, we show the redshift evolution of the same fields along one given direction of the corresponding fields.

\begin{figure*}
	\includegraphics[width=1.05\textwidth]{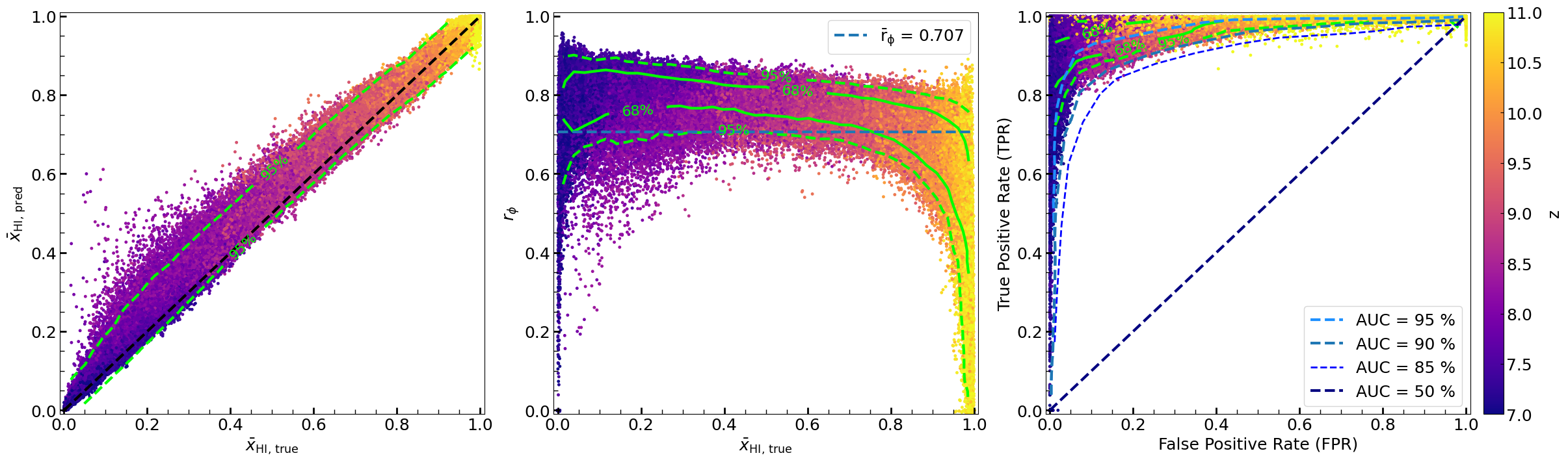}\vskip-2mm
	\caption{Statistical analysis of the predicted binary maps for the testing dataset. Each point indicates an image at a given redshift in the colour bar. \textit{Left Panel}: correlation plot between the ground truth volume average neutral fraction, $\overline{x}_{\rm HI,\,true}$, against the predicted, $\overline{x}_{\rm HI,\,pred}$. \textit{Right Panel}: Matthew correlation coefficient $r_{\phi}$ against global volume-averaged neutral fraction. The dashed blue line indicates the redshift averaged $r_{\phi}$. Here, solid green lines indicate the 68 per cent ($1\sigma$) and dashed green lines the 95 per cent ($2\sigma$) data contour. \textit{Right Panel}: Receiver Operating Characteristic curve for the same dataset. The dashed line of different blue shades indicates the percentage of reliability of the prediction.}
	\label{fig:mcc}
\end{figure*}

First, when we compare the bottom right panel in \autoref{fig:train_example} with the top right panel in \autoref{fig:vis_eval}, we can notice that the pre-processing step drastically reduces the signal from $\delta T_b\sim\pm 10^5\,\rm mK$ to just an observed differential brightness of few tens $\delta T_b\sim\pm 40\,\rm mK$. Nevertheless, some of the foreground contamination is still visible. For instance, in \autoref{fig:vis_eval} top left panel, we can see that across a few frequency bands at $z\approx 10.8$ presents an anomalous feature. Moreover, we can see that foreground residual is still present between $7\leq z \leq 8.2$. This signal excess is self-evident in the per-pixel uncertainty for the same redshift range. Some frequency bands are saturated with considerable uncertainty $\sigma_{\rm std}\sim 0.3$. This is because the foreground component is correlated along the frequency direction and is primarily diffused over large angular scales. The foreground residuals thus observe extended features along the $z$ direction over multiple adjacent frequency channels. From the redshift evolution of the predicted binary field (left middle panel), we notice that the network can either falsely detect bubbles when most of the lightcone is still highly neutral, $z\geq 9.5$, or completely miss ionised bubbles that are entirely surrounded by neutral hydrogen. In both cases, the mislabelling is limited to bubbles with sizes close to or smaller than the interferometric smoothing scale, $\Delta x\sim9\,\rm Mpc$, as the network confuses structures with small-scale noise fluctuations. Thus posing a hard limit on the possibility of measuring and detecting the smallest {\sc HII} bubble close to the instrument resolution. We discuss this further in \S\ref{sec:ionbubble}. This limitation is visible from the recovered binary field at redshift $z=8.24$ (middle right panel). Here, the detection of the bubbles at $ 180\,\rm Mpc \leq x \leq 210\,\rm Mpc$ is entirely missed. We observe the same outcome for the island of neutral hydrogen at coordinates $(x,y)\approx (75, 75)\,\rm Mpc$. These erroneous findings are associated with a moderate to high uncertainty $\sigma_{\rm str} \geq 0.2$. As we mentioned above, the per-pixel uncertainty shows that at the early stage of reionization, $z>9$, most of the uncertainty is either situated around small {\sc HII} volumes, $V \leq (10^3\,{\rm Mpc})^3$, or at the border between neutral and ionised regions. On the other hand, at the late stages, $z<8.2$, high uncertainty is mostly located in the vast, interconnected ionised IGM. 

In \autoref{fig:mcc}, we show three statistical analyses for the entire \textit{random testing set}. In the left panel, we show the correlation plot between the true global averaged neutral fraction $\bar{x}_{\rm HI,true}$ against the predicted $\bar{x}_{\rm HI,pred}$. The dashed green line indicates the 95 per cent data contour, corresponding to a $2\sigma$ difference from the ground truth. The $2\sigma$ contour clearly shows a deviation on the left-hand side of the black dashed line (perfect correlation), indicating that the predicted images tend to be considered more neutral than they should be. This trend is more visible at lower redshift $z<8.5$ ($\bar{x}_{\rm HI,true}<0.4$) as more points reside outside of the $95\%$ percentile. This behaviour can be motivated by the presence of residuals from the foreground that the PCA process could not remove. As we mention in \S\ref{sec:pca}, we consider the first four components to contain most foreground information. These components are most representative at higher frequency as the foreground amplitude increases inversely proportional to redshift, \autoref{eq:foreground_powerspect}. Therefore, for tomographic data with a wide redshift range, the decomposition can under-represent foreground contamination at lower redshift, resulting in more residuals when we reconstruct the image from the remaining components at the corresponding redshift slices. This effect is visible in the uncertainty map in \autoref{fig:vis_eval}. 

In \autoref{fig:mcc}, middle panel, we show the correlation coefficient against the same quantity as before, $x_{\rm HI,true}$. Each point corresponds to an image at a redshift indicated by the colour bar. We add the 68 per cent data contour (solid line) on this panel, corresponding to a $1\sigma$ difference from the ground truth. We first noticed that we obtain a global accuracy that is approximately $15\%$ lower, $\bar{r}_{\rm\phi}=0.71$, compared to our previous work in \citetalias{Bianco2021segunet}. This lower score with the same network structure and architecture is justified because any signal extrapolation in foreground contamination is extremely arduous compared to forecasting in the presence of just telescope systematic noise. Moreover, as we stated before, we notice that at lower redshift $z<8.5$ ($\bar{x}_{\rm HI,true}<0.4$), a sizable portion of the redshift slices have a difference larger than $2\sigma$. This behaviour is also evident from the increase of the uncertainty map in \autoref{fig:vis_eval} for images at $z<8.5$.

Lastly, in \autoref{fig:mcc}, right panel, we show the correlation between the true positive rate ($TPR$), also known as sensitivity, and the false positive rate ($FPR$), also known as non-specificity, on the \textit{random testing set}. In our case, these quantities indicate the percentage of pixels correctly labelled as neutral and the fraction of pixels mislabelled as ionized, respectively. This plot is known as the Receiver Operating Characteristic (ROC) curve, and it is a standard analysis in classification problems as it gives an intuitive overall performance of the method. The results from our network show that most of the realizations with redshift range $z\in[7.5,\,10]$ are located in the top-left corner, representing the ideal performance or perfect classification. This indicates that most binary maps have high sensitivity and specificity, i.e., neutral and ionised regions are correctly identified. Data points close to the diagonal line indicate that the method performance is not much better than a random classifier. In our case, this is true for the values at the extreme of the redshift range. The data points on the top-right corner have high sensitivity but low specificity, meaning that the network labels correctly neutral regions, from \autoref{eq:fpr_tpr}, left metric, $FN\ll TP$, while misclassifying most of the ionized pixels as neutral, $TN\ll FP$. This is the case for images with $z>10$; however, at this redshift, the images are mostly neutral; thus, the incorrect detection is limited to a few pixels of the image. The data point in the bottom-left represents the opposite situation where the network has high specificity but low sensitivity. This scenario indicates that the model is not able to differentiate well between neutral and ionized instances, from \autoref{eq:fpr_tpr}, right function, $TP\ll FN$ and $FP\ll TN$. We see the opposite trend as in the previous case, where images with $ z\sim 7$ occupy this instance. Another important quantity derivable from the ROC curve is the area under the curve ($AUC$). This quantity gives an overall evaluation of the classification method. In \autoref{fig:mcc}, right panel, we overplot four curves that represent different $AUC$ scores. In our case, we can see that the network performs well as the \textit{random testing set} points are mostly located above the $85\%$ line and are well centred around the $AUC=95\%$.

\subsection{Sensitivity to the Choice of Pre-processing Method}\label{sec:other_preproc} 
\begin{figure*}
	\includegraphics[width=1.01\textwidth]{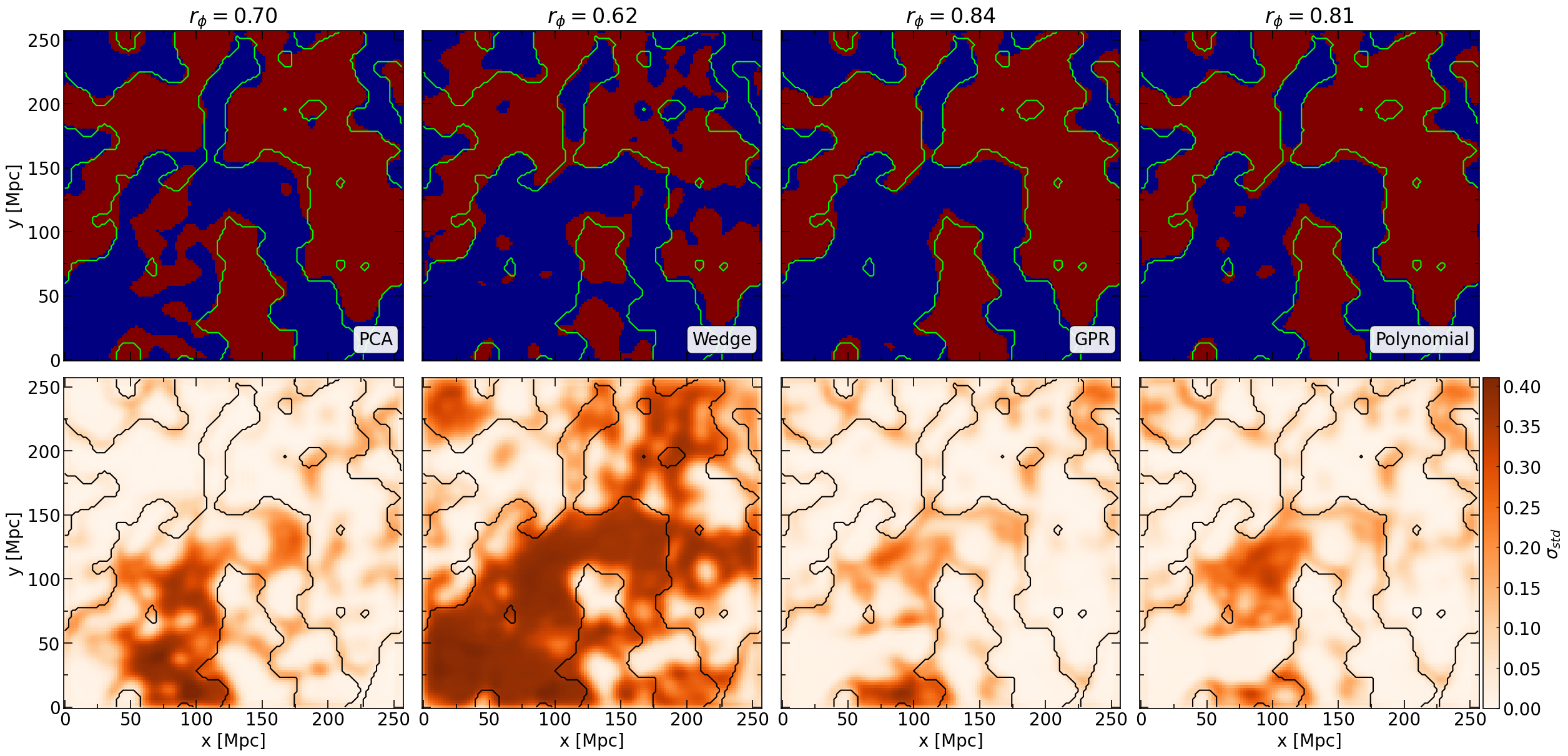}\vskip-2mm
	\caption{Comparison of the recovered binary field from different foreground mitigation pre-processes. We have PCA, wedge removal, GPR, and polynomial fitting from left to right. \textit{Top panels}: a visual example of the recovered binary map at redshift $z=8.24$ after the mentioned pre-processing step. The red/blue indicates the predicted neutral/ionized regions, while the green contour indicates the ground truth. \textit{Bottom panels}: the corresponding per-pixel uncertainty map derived by \texttt{SegU-Net v2}. The orange indicates the intensity of the uncertainty, defined as a general standard deviation. The title includes the resulting $r_{\phi}$ at this redshift.}
	\label{fig:preprocess_comparison}
\end{figure*}

We trained \texttt{SegU-Net v2} on the signal that is pre-processed using the PCA method. Therefore, it is vital to investigate how sensitive the trained model is to the pre-processing method used to mitigate foreground. Here, we test \texttt{SegU-Net v2} on the foreground mitigation processes we presented in \S\ref{sec:frg_mitigation}. We cannot use the entire lightcone as the GPR module currently available has been validated only for a bandwidth of 20 MHz. From the entire lightcone, we use three sub-volume centred at redshift $z_c=7.68$, $8.24$ and $8.97$ with frequency size of $20\,\rm MHz$, corresponding to $172$, $181$ and $186$ redshifts bins from $z\in[7.19,\,8.24]$, $[7.68,\,8.88]$ and $[8.31,\,9.72]$, respectively. The volume average neutral fraction of these sub-volumes is $\overline{x}_{\rm HI}\simeq 0.25$, $0.50$ and $0.75$, corresponding to the late, middle and early stages of reionization, respectively. 

\begin{table*}
	\centering
	\caption{Result summary of the predicted binary field for the tested pre-processing step on the three lightcone sub-volume at representative 
    stages of reionization.}
	\label{tab:summary_results}
    \def\arraystretch{1.25}
	\begin{tabular}{lcccccccccc}
	\hline 
    $z_c$  & pre-process  & $r_\phi(z_c)$ & Accuracy & Precision & IoU & $TPR$ [$\%$] & $FPR$ [$\%$] & $\overline{r}_\phi$ & $\overline{x}_{\rm HI}$ & $\overline{R}_C\,\rm[cMpc]$ \\ \hline
    $7.68$ & Ground Truth & -      & -      & -      & -      & -      & -      & -      & $\bf 0.24$     & $\bf 19.89$             \\
           & all z PCA    & $0.78$ & $0.94$ & $0.81$ & $0.67$ & $83.12$& $5.98$ & $0.82$ & $0.26\pm0.12$  & $21.62^{+4.34}_{-3.90}$ \\ 
           & PCA          & $0.75$ & $0.89$ & $0.81$ & $0.70$ & $84.08$& $8.32$ & $0.73$ & $0.26\pm0.15$  & $17.96^{+8.66}_{-4.66}$ \\
           & Wedge        & $0.55$ & $0.80$ & $0.65$ & $0.20$ & $52.56$& $49.82$ & $0.28$ & $0.07\pm0.12$  & $11.96^{+9.46}_{-2.54}$ \\
           & GPR          & $0.77$ & $0.90$ & $0.82$ & $0.73$ & $86.22$& $7.97$ & $0.77$ & $0.28\pm0.14$  & $19.75^{+6.93}_{-5.03}$ \\
           & Polynomial   & $0.75$ & $0.89$ & $0.82$ & $0.70$ & $83.81$& $8.09$ & $0.76$ & $0.27\pm0.15$  & $19.17^{+7.84}_{-5.18}$ \\\hline
    $8.24$ & Ground Truth & -      & -      & -      & -      & -      & -      & -      & $\bf 0.45$     & $\bf 29.54$             \\
           & all z PCA    & $0.84$ & $0.91$ & $0.86$ & $0.72$ & $90.60$& $5.32$ & $0.80$ & $0.48\pm0.07$  & $31.37^{+3.09}_{-3.93}$ \\
           & PCA          & $0.70$ & $0.85$ & $0.81$ & $0.75$ & $91.12$& $21.48$ & $0.69$ & $0.49\pm0.11$  & $27.65^{+9.13}_{-6.12}$ \\
           & Wedge        & $0.62$ & $0.64$ & $0.65$ & $0.22$ & $74.95$& $45.43$ & $0.22$ & $0.16\pm0.13$  & $15.20^{+24.13}_{-6.18}$\\
           & GPR          & $0.84$ & $0.92$ & $0.91$ & $0.85$ & $93.02$& $9.44$ & $0.75$ & $0.48\pm0.09$  & $29.14^{+5.26}_{-4.89}$ \\
           & Polynomial   & $0.81$ & $0.91$ & $0.89$ & $0.83$ & $92.18$& $11.01$ & $0.74$ & $0.49\pm0.10$  & $29.21^{+5.83}_{-5.21}$ \\\hline
    $8.97$ & Ground Truth & -      & -      & -      & -      & -      & -      & -      & $\bf 0.72$     & $\bf 49.09$             \\
           & all z PCA    & $0.78$ & $0.92$ & $0.93$ & $0.85$ & $93.43$& $15.52$& $0.76$ & $0.74\pm0.29$  & $48.57^{+5.93}_{-6.36}$ \\
           & PCA          & $0.72$ & $0.88$ & $0.90$ & $0.85$ & $93.80$& $23.75$ & $0.68$ & $0.75\pm0.33$  & $46,06^{+9.47}_{-8.74}$ \\
           & Wedge        & $0.53$ & $0.51$ & $0.76$ & $0.37$ & $70.96$& $77.96$ & $0.19$ & $0.38\pm0.11$  & $28.57^{+11.46}_{-8.54}$\\
           & GPR          & $0.75$ & $0.90$ & $0.91$ & $0.86$ & $94.53$& $22.10$ & $0.72$ & $0.74\pm0.28$  & $46.64^{+7.21}_{-7.52}$ \\
           & Polynomial   & $0.74$ & $0.89$ & $0.90$ & $0.86$ & $94.53$& $22.78$ & $0.72$ & $0.74\pm0.29$ & $47.24^{+7.07}_{-7.81}$ \\ \hline
	\end{tabular}
\end{table*}

We then apply four different foreground mitigation pre-processing steps to each sub-volume: PCA, Wedge Remove, GPR and Polynomial fitting. From the residual volumes, we predict the neutral/ionised regions from the trained \texttt{SegU-Net v2}, with PCA, pre-processing step as presented in \S\ref{sec:evaluation}. By applying different foreground mitigation processes, we can quantify the robustness and adaptability of our trained network. 
\begin{figure*}
	\includegraphics[width=\textwidth]{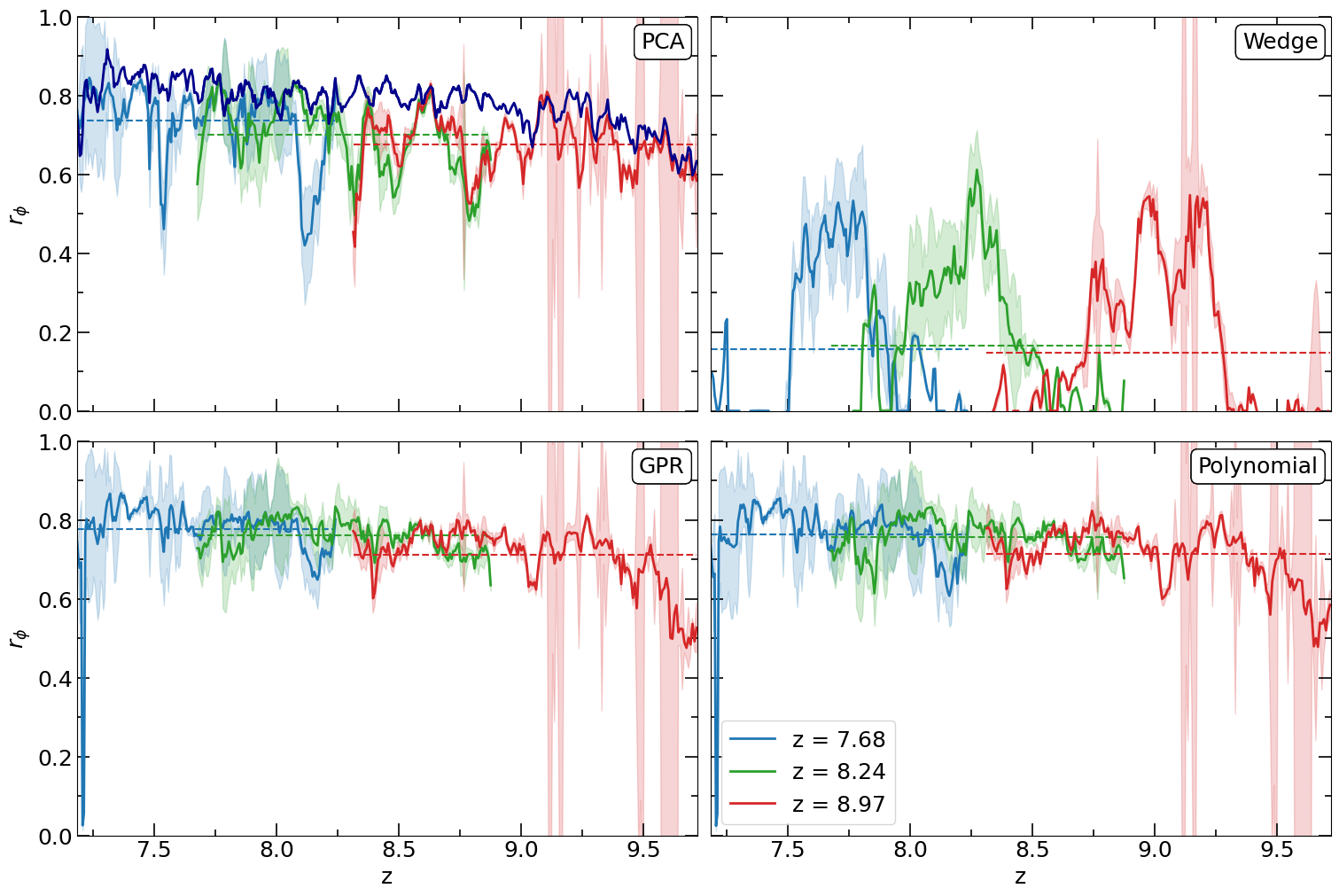}\vskip-3mm
	\caption{Redshift evolution of the $\rm r_{\phi}$ correlation coefficient for the different tested pre-processing step. Each panel shows the result on three lightcone sub-volumes centred at $z_c=7.68$ (blue), $8.24$ (green) and $8.97$ (red) with a $\pm10\,\rm MHz$ frequency depth. These redshifts correspond to the late, middle and early stages of reionization, respectively. Solid lines indicate the $r_{\phi}$ coefficient for the predicted binary maps. Shadow areas indicate the error due to the uncertainty map. Horizontal dashed lines indicate the redshift averaged $\overline{r}_{\phi}$ coefficient. For the case of PCA, we plot the decomposition executed on the full redshift range (dark blue) as a reference.}\label{fig:preprocess_comparison_redshift}
\end{figure*}
\subsubsection{Visual Evaluation}
We visually compare the middle stage of reionization sub-volume for the four cases in \autoref{fig:preprocess_comparison}. From the left to right column, we have PCA, Wedge Remove, GPR and Polynomial fitting, respectively. The top panels visually compare an image at the sub-volume central redshift $z_c=8.24$ for the different pre-processes. In the bottom panels, we show the corresponding uncertainty map from the \texttt{SegU-Net v2}. 
We notice that for the case of the fiducial simulation, the Polynomial fitting and GPR pre-processing obtain similar results with correlation $r_{\phi}(z_c)=0.81$ and $r(z_c)_{\phi}=0.84$, respectively. The former case appears to overestimate the extent of the neutral regions (see at position $(x,y)\simeq(75,125)\,\rm Mpc$) as well as falsely detecting the presence of isolated neutral island in the vast ionised region, for instance, see around $(x,y)\sim(75,100)\,\rm Mpc$. The PCA obtains approximately $10\%$ less accuracy, $r_{\phi}(z_c)=0.70$, its limitation comes forth when predicting the vast ionised region (see at position $50\,{\rm Mpc}\leq x\leq 125\,{\rm Mpc}$ and $75\,{\rm Mpc}\leq y\leq 125\,{\rm Mpc}$) as the network is over-predicting the presence of an interconnected neutral hydrogen region. Wedge Remove method has the lowest performance, with $r_{\phi}(z_c)=0.62$. In this example, the pre-process forecasts an excess of neutral hydrogen outside the ground truth. On the other hand, this method underestimates its presence within the extensive neutral cloud. In \autoref{tab:summary_results} third column, we show the resulting $r_\phi(z_c)$ for each pre-process.

Among the methods presented, the Wedge Remove method appears to be the least efficient for \texttt{SegU-Net v2}. The uncertainty map in \autoref{fig:preprocess_comparison} shows that the Wedge Remove method has high incertitude in the vast interconnected {\sc H~ii} regions, for $x\in[0,\,125]\,\rm Mpc$ and $y\in[0,\,150]\,\rm Mpc$, as well as between nearby {\sc H~i} regions, for instance at $(x,\,y)\simeq(120, 160)\,\rm Mpc$. The presence of a higher foreground residual compared to the other methods (visible in the same region in \autoref{fig:example_comparison}) indicates that lower performance is attributed to a harsh and perhaps undisclosed subtraction that does not aim at portraying the foreground contamination but rather removes its contribution. Overall, the GPR method, followed by PCA decomposition, appears to give an advantage compared to the other pre-processing. At the same time, all the cases fail to detect ionised or neutral regions of sizes close to the interferometric smoothing scale, $\Delta x \simeq 9\,\rm Mpc$.

\subsubsection{Redshift Evolution}
In \autoref{fig:preprocess_comparison_redshift}, we show the redshift evolution of the Matthew correlation coefficient $r_{\phi}$ for the four different methods. On each panel, we show the results from the early ($z_c=8.97$, in red), middle ($z_c=8.24$, in green) and late ($z_c=7.68$, in blue) stage of reionization sub-volumes with the corresponding error bar represented by the shadow area. The horizontal dashed line denotes the redshift averaged correlation coefficient, $\overline{r}_{\phi}$. In \autoref{tab:summary_results} fourth column, we show the resulting $\overline{r}_\phi$ for each sub-volume and sub-volume. Based on this quantity, we notice that the ranking goes by the GPR method with $\overline{r}_{\phi}=0.71$ at $z_C=7.68$, $0.67$ at $z_C=8.24$ and $0.63$ at $z_C=8.97$, followed by the PCA with $\overline{r}_{\phi}=0.68$, $0.67$ and $0.62$, respectively. Polynomial fitting follows with $\overline{r}_{\phi}=0.65$, $0.62$ and $0.60$, while Wedge Remove follows with $\overline{r}_{\phi}=0.18$, $0.19$ and $0.15$, respectively. An important remark: in this comparison, we limit the PCA decomposition to the sub-volumes redshift bins ($172$, $181$ and $186$), and it is performing slightly worse when compared to the same results in the previous section on the $552$ redshift bins. Therefore, we attribute the performance decrease to the reduced number of redshift bins that directly lower the number of orthogonal components with which the data are represented. For the case of PCA in \autoref{fig:preprocess_comparison_redshift}, we plot on the same panel the performance of the PCA decomposition on the $552$ redshifts (dark blue line). Here, we can notice how the redshift averaged correlation coefficient is substantially higher, $\overline{r}_{\phi}=0.82$ at $z_C=7.68$, $0.80$ at $z_C=8.24$ and $0.76$ at $z_C=8.97$, hence indicating that the PCA pre-process is preferred if we have at our disposal a tomographic dataset with an extended redshift range. The sharp increase at $z\simeq 8.76$, the sudden increase at $z \geq 9$ and the constant broadening for $z\leq 8.1$ of the uncertainty error in \autoref{fig:preprocess_comparison_redshift} indicates that the PCA, GPR and Polynomial fitting are sensible to the evolution and distinctiveness of the same structures in the data.

Moreover, all processes, except for PCA, show a slight decrease in accuracy close to the redshift extremities values of the sub-volume. The Wedge removal efficiently helps recover the binary maps only for the selected sub-volume central part, close to the central redshift. While the accuracy decreases rapidly toward the edges as the foreground removal becomes inefficient, in our simplified version of the wedge removal code, we do not include the sliding trough process. See \S\ref{sec:wedge}. Therefore, a comparison between the Wedge removal and the other pre-processing should be strictly limited to the sub-volume central part.

\subsubsection{Recovered Neutral Island Size Distribution}\label{sec:isd}
\begin{figure*}
	\includegraphics[width=\textwidth]{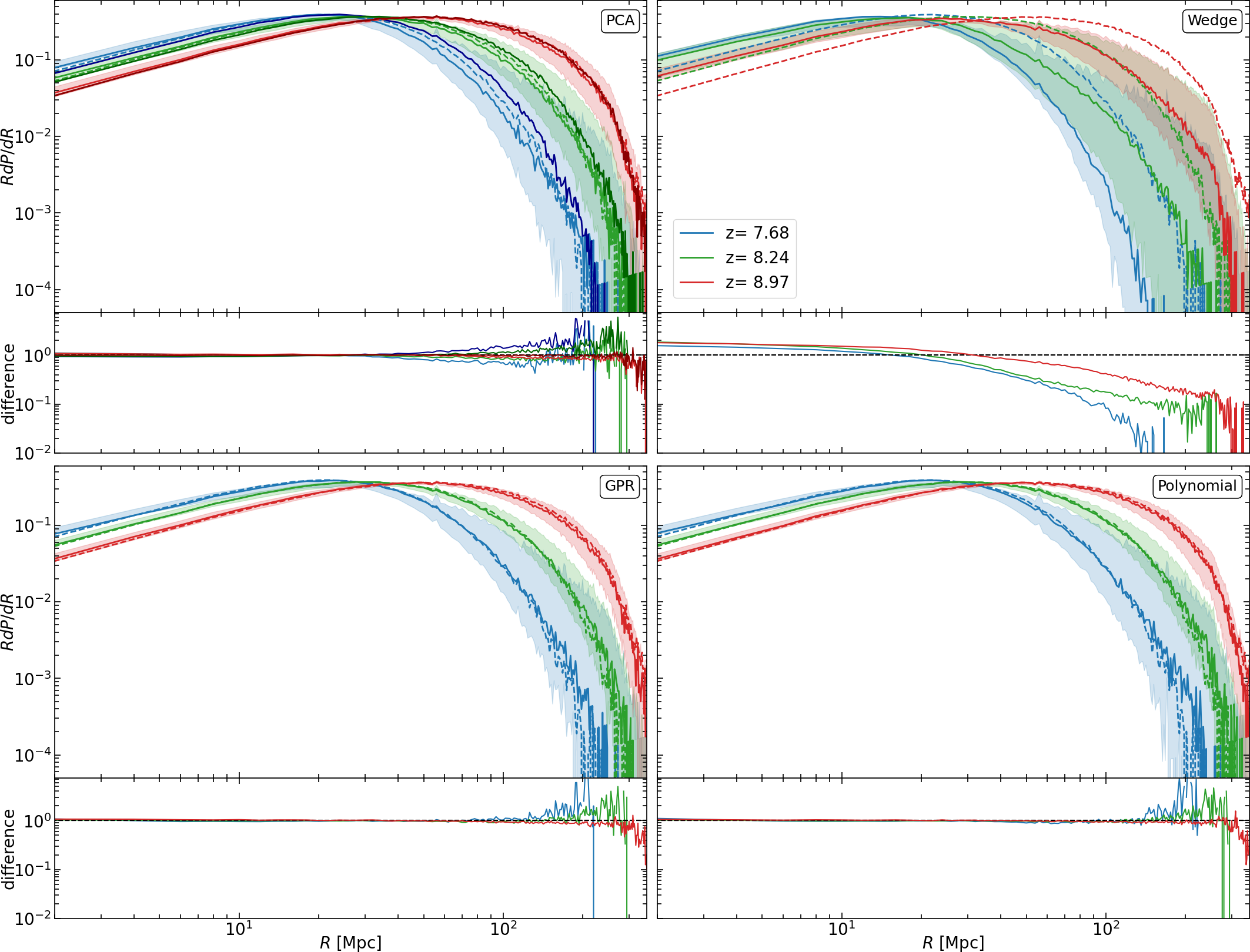}\vskip-3mm
	\caption{Island size distribution for the different pre-processing steps. Each panel shows the predicted size distribution $R\,dP/dR$ (top section) and the difference to the ground truth (bottom section). The colours indicate the lightcone sub-volume at the late  ($z_c=7.68$, blue), middle ($z_c=8.24$, green) and early ($z_c=8.97$, red) stage of reionization. The results from the neutral regions in the predicted fields are shown with solid lines and the ground truth with dashed lines. For the case of PCA, we plot as a reference the predicted size distribution with a dot-dashed line.}
	\label{fig:preprocess_comparison_isd}
\end{figure*}

In \autoref{fig:preprocess_comparison_isd}, we compare the neutral island size distribution (ISD) derived from the {\sc Hi} binary field predicted with the different pre-processing methods presented in \S\ref{sec:frg_mitigation}. We employ the Mean-Free Path \citep[MFP;][]{Mesinger2007EfficientReionization} method to derive the probability density distribution ($RdP/dR$) of the neutral region sizes or radius $R$. This size distribution measures the topological evolution of the reionization process \citep{Friedrich2011TopologyReionization, Giri2018BubbleTomography}. See \citet{Giri2019NeutralTomography} for a detailed study of ISDs during reionization.

In \autoref{fig:preprocess_comparison_isd}, each panel shows the predicted ISD (solid line) for three sub-volumes centred at redshift $z_c=7.68$ (blue), $8.24$ (green) and $8.97$ (red) against the ground truth ISD (dashed line). In the bottom part of each panel, we show the difference with the ground truth. Similarly to before, in the case of PCA, the estimated distribution with PCA decomposition on the full redshift range, from $7$ to $11$, is shown with a darker colour. We show the uncertainty error on the predicted ISD with a shadow area of the same colour. The GPR method and the polynomial fitting from neutral island distribution analysis appear to be the best fit. Differences are visible only at a large scale, $R\geq 100\,\rm Mpc$, with a factor $\sim 3$ larger for the early and middle reionisation sub-volume stage. The only noticeable difference for the early stage sub-volume is for the extremely large sizes, $R\approx300\,\rm Mpc$. 
The results from the training pre-processing (darker colour) predict an ISD consistently shifted toward a larger scale for the case of $z_c=7.68$ and $8.24$. Deviations from the ground truth start to be visible for scale $R\geq40\,{\rm Mpc}$ and $R\geq80\,\rm Mpc$ with differences from up to a factor of $\sim 2$ and a maximum of $5$ at $R\approx 200\,\rm Mpc$. On the other hand, for the case of the sub-volume centred at $z_c=8.97$, the predicted ISD shows no virtual difference. These results confirm what we concluded in \S\ref{sec:evaluation}, with the analysis from \autoref{fig:mcc} (left panel). The PCA performed on the sub-volume redshift range shows the same factorial difference but with an opposite behaviour. Differences are more prominent for the late stage of reionisation sub-volume and get gradually better at the early stage. In this analysis, the Wedge method fails to depict the {\sc Hi} distribution for all the sub-volumes. For small neutral regions, $R\leq 20\,\rm Mpc$, the predicted distribution is a factor $2$ larger, while for larger sizes, the distribution can be severely underestimated, with $RdP/dR$ two orders of magnitude smaller than the ground truth distribution. This performance is an indication that with the Wedge pre-processing, \texttt{SegU-Net v2} is struggling to connect large neutral regions due to the missing 21-cm signal lying in the \textit{foreground wedge} region that has been removed along with the foreground.

From the probability density distribution $RdP/dR$, we can estimate the mean radius of the neutral islands at a given redshift, defined as 
\begin{equation}
\overline{R}_C(z) = \int^{\infty}_{R_{\rm min}} R\,\frac{dP}{dR}(z)\,dR\,.
\end{equation}
In our case, we set the lower limit to the intrinsic resolution of our simulation $R_{\rm min} = 2\,\rm cMpc$. In \autoref{tab:summary_results}, rightmost column, we list this quantity derived from the predicted binary field with the different pre-process. The ground truth average radius is $\overline{R}_C=19.89\,\rm cMpc$ for the sub-volume centered at $z_c=7.68$, $\overline{R}_C=29.54\,\rm cMpc$ for $z_c=8.24$ and $\overline{R}_C=49.09\,\rm cMpc$ for $z_c=8.97$. Based on this quantity, we notice that the GPR method and Polynomial fitting produce a better prediction for the late and middle EoR sub-volumes, with a difference to the ground truth below the $\rm cMpc$, while for the early stage scenario, they tend to underestimate of a few $\rm cMpc$. In the case of both PCA decompositions, the predicted quantity differs by a few $\rm cMpc$ in excess and deficit, respectively. This trend is also visible from the predicted ISD, as PCA shows a systematic underestimation, while the same decomposition on the entire redshift range shows an overestimation for the same scale, $R\geq 30\,\rm cMpc$. Considering the uncertainty, the wedge method seems to work reasonably well only for the late stage of reionization. However, for this scenario, the predicted ISD does not match. At late stages, the Wedge Removal prediction of $\overline{R}_C$ can not be trusted, as this quantity differs substantially.

 \subsection{Relation between ionised volume and total ionising photons}\label{sec:ionbubble}
\cite{Zackrisson2020Bubble6} illustrated the possibility of employing SKA-Low tomographic data as a foreshadowing method to identify the region of interest for future and ongoing experiments that aim to observe galaxy formation in the early Universe, such as the JWST, Euclid and Nancy Grace Roman Space Telescope \citep[e.g.][]{Beardsley2015, Geil2017}. This work demonstrated that there is a simple relation between the volume of isolated {\sc Hi}I bubbles, $\rm V_{ion}$, and the grand total of ionising photons, $\rm N_{\gamma,\, tot}$, produced by the primordial sources within the same ionised region.
\begin{figure}
	\includegraphics[width=\columnwidth]{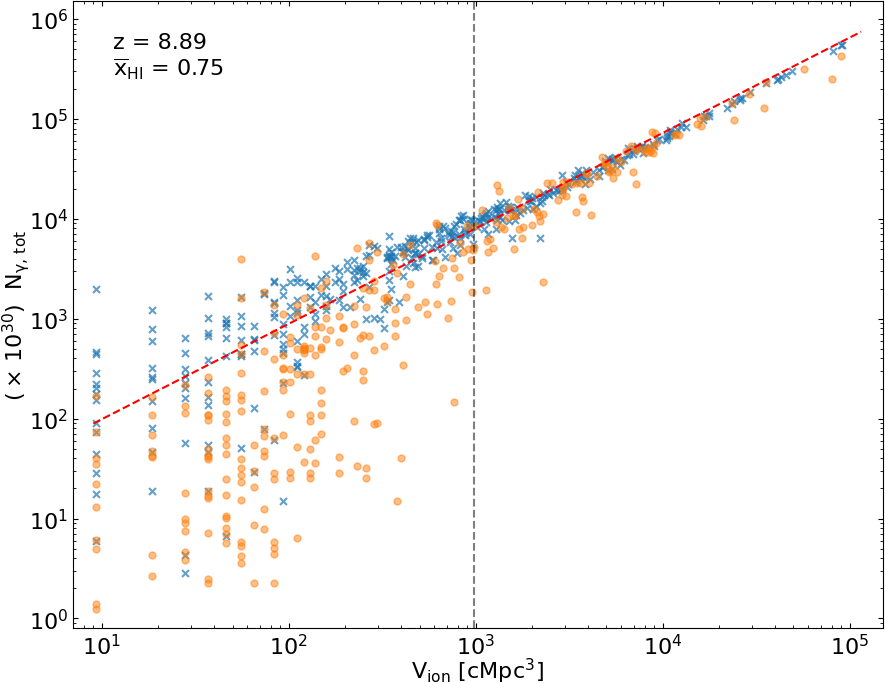}\vskip-2mm
	\caption{Relation between the volume of ionised region versus the grand total of ionising photons within the same region. For a coeval cube at redshift $z=9$ ($\rm\overline{x}_{HI}=0.75$) and box size of $\rm L_{box}\approx348\,cMpc$. Relation derived from the ground truth is represented with blue cross data, while orange circle points are derived from \texttt{SegU-Net} prediction. The dashed red line corresponds to the linear fit of the ground truth data points. The vertical line indicates the $\rm 2\,km$ baseline smoothed resolution.}
	\label{fig:volume_relation}
\end{figure}
Although we are overlooking relevant instrumental effects (e.g. incomplete uv-coverage, absence of gain error, beam effect and more), we assume that our framework, described in \S\ref{sec:mock_obs}, produces realistic enough mock observation to demonstrate the challenge of identifying and measuring the sizes of such bubbles and its derived relation.

For this analysis, we require the mass and the position of the sources within the ionised bubbles. Therefore, we decided to use a simulation run with the \texttt{C$^2$Ray} radiative transfer code \citep{Mellema2006CRadiation}. In \citetalias{Bianco2021segunet}, we demonstrated how \texttt{SegU-Net} works reasonably well on simulations other than those employed for the training and validation. Moreover, recent works demonstrated the limitations of U-Net when cross-validating different cosmological models \citep{Chen2023stability}. Here, we employ the obtained ionised hydrogen and density coeval cubes to calculate the 21-cm differential brightness with \autoref{eq:dTb} and follow the mock observation procedure explained in \autoref{sec:mock_obs}. We consider the third axis the frequency direction to create the corresponding network input and target. We use one realisation of the simulated coeval cube at redshift $\rm z=8.89$ with box and mesh size of $\rm 348\, cMpc$ and $250$, respectively. We interpolate the $\rm 250$ mesh grid into a $\rm 166$ grid per side to a corresponding intrinsic resolution similar to our $\rm\Delta x = 2.09\, Mpc$ dataset. One of the inputs of the \texttt{C$^2$Ray} code is the cumulative halo mass smoothed into the mesh grid. In this way, we can associate an ionised bubble to the sources within the same region by converting the total halo distribution mass $M_{\rm h, tot}$ to the total ionising photon produced $\rm N_{\rm\gamma, tot}=f_{\gamma}\,\Omega_{m}/\Omega_b\, M_{\rm h, tot}$. We refer the reader to \cite{Iliev2006SimulatingReionization, Iliev2012CanSources} and \cite{Dixon2016TheReionization} for further reading on the halo source model.

Though \texttt{SegU-Net v2} is not trained on simulations produced with \texttt{C$^2$Ray}, we still find that the ionised regions are accurately identified. This analysis shows that the trained model is quite general\footnote{We should note that we have not tested the framework on radiative transfer hydrodynamical simulations due to the unavailability of models with box lengths exceeding 200 Mpc, which is essential for studying the 21-cm signal \citep[e.g.,][]{giri2023suppressing}.} and, therefore, capable of finding physical features in real observations. In \autoref{fig:volume_relation}, we show the relation between $\rm V_{ion}$ and $\rm N_{\gamma,\, tot}$ derived from the simulation data (blue crosses) and the predicted binary maps (orange points). We notice that \texttt{SegU-Net v2} is failing to correctly quantify the number of ionising photons for volumes $\rm V_{ion}\lesssim(10\,cMpc)^3$, vertical black dash line. This limitation corresponds to the $\rm 2\,km$ interferometric smoothing scale we apply in our mock observation pipeline. At $z=8.89$, the Gaussian kernel has an angular scale of $\Delta\theta\approx 3.57\rm\,arcmin$, corresponding to a comoving size of $\rm 9.9\,cMpc$. This limitation is also consistent with the results in \autoref{fig:mcc}, where the correlation between prediction and ground truth slowly decreases, $r_{\phi}\leq80\%$, for higher redshift, $\rm z\geq 9$.

\section{Discussion \& Conclusions}\label{chap:discussion}
With this work, we improved our previous effort in \citetalias{Bianco2021segunet} and updated our deep learning framework, \texttt{SegU-Net v2}, for the identification of neutral and ionized regions in realistic 21-cm mock observation expected from SKA-Low. One of the advantages of our network is the possibility to provide per-pixel uncertainty maps on its predictions. In \S\ref{sec:mock_obs}, we introduced our extended mock observation pipeline by including synchrotron Galactic foreground contamination, presented in \S\ref{sec:frg}. Additionally, we performed machine learning hyper-parameter optimisation. We show the best-performing hyperparameters setup we analysed in \S\autoref{app:hyperpar}.

In this work, we combine our network with a foreground mitigation method that pre-processes the input data and reduces, in part, the foreground contribution. We trained \texttt{SegU-Net v2} on $10,000$ lightcones with $552$ redshift slices from $z=7$ to $11$ pre-processed with PCA on $4$ components for the full redshift range. We chose this pre-processing method as it is the most commonly used method for foreground contamination and provides fast and efficient mitigation. In \S\ref{sec:evaluation}, the analysis on a random sample dataset, composed of $300$ lightcone with the same redshift extent and bins, shows that the updated version of our network works well, with an average correlation of $71\%$, on 21-cm images contaminated and pre-processed by a foreground contamination method. This level of accuracy is almost $\sim 20\%$ less than our previous results and is attributed to the added complexity due to the presence of the Galactic foreground. We show that \texttt{SegU-Net v2} recovered binary fields that tend to be considered more neutral at $z\leq 8.5$. We attribute this to the under-subtraction of the PCA pre-processing method employed during training. This trend is confirmed by the increase of the uncertainty map for the same redshift extent that saturates entire frequency channels (see the bottom panel in \autoref{fig:vis_eval}).

In \S\ref{sec:other_preproc}, we compared the binary maps predicted with \texttt{SegU-Net v2} on different pre-processing foreground mitigation and one avoidance method. We consider three sub-volume of the fiducial simulation with frequency width $\Delta\nu=\pm10\,\rm MHz$ centred at redshift $z_c=7.68$, $8.24$ and $8.97$, representing a late, middle and early stage of reionisation. In this work, we consider PCA decomposition (\S\ref{sec:pca}), Wedge removal (\S\ref{sec:wedge}), Gaussian Process Regression (\S\ref{sec:gpr}) and Polynomial fitting (\S\ref{sec:poly}). 
We demonstrated that \texttt{SegU-Net v2} is able to recover {\sc Hi} regions with varying accuracy for all the pre-processing methods we tested. In our case, the network is able to generalize enough and work with the same level of accuracy as the training case on pre-processing methods that were not employed during its training (see summary statistics in \autoref{tab:summary_results}). Moreover, in \S\ref{sec:isd}, we study the island size distribution (ISD) of the predicted binary maps. GPR and Polynomial fitting work better in recovering the ISDs, as well as the average distribution size $R_C$ of neutral regions, than the two cases of the PCA pre-processing (applied on the full redshift range and the sub-volume redshift range). 

Therefore, we can conclude that \texttt{SegU-Net v2} is the pre-processing method agnostic, providing accurate predictions independent of the pre-processing method, as long as the foreground mitigation provides reasonable residual images of the original 21-cm signal. Another conclusion is that PCA decomposition on lightcone data with a wide redshift range, e.g. frequency depth of the order of $60\,\rm MHz$ or larger, is to be preferred. In the case of smaller available sub-volumes, with frequency depth between $20\,{\rm MHz}$ and $30\,{\rm MHz}$, other methods such as GPR or Polynomial fitting are to be preferred as they provide better prediction when compared to PCA on the same redshift range.

Finally, we provided a concrete use case of \texttt{SegU-Net v2} in the context of 21-cm SKA-Low tomographic observation. Previous work demonstrated that a linear relation could be derived between the size of the ionised volume and the grand total number of ionising photons produced by the hosted source. In \S\ref{sec:ionbubble}, we demonstrated that our network could recover with precision the linear relation for ionised volumes that are resolved. Here, we stipulate the limited resolution of the SKA-Low layout by the interferometric smoothing scale for the maximum baseline of $B=2\,\rm km$, which corresponds to an angular scale of approximately $ 3.57\, \rm arcmin$ at redshift $z=8.89$, corresponding to an early stage of reionisation scenario, $\bar{x}_{\rm HI} = 0.75$. 

The current version of \texttt{SegU-Net v2} is trained using semi-numerical simulations, known for their non-conservation of photons \citep[e.g.,][]{hutter2018accuracy,choudhury2018photon}. This discrepancy arises when the number of photons emitted by the sources does not match the number of IGM ionisations. However, it is important to highlight that \texttt{SegU-Net v2} does not exhibit sensitivity to the model linking the sources and sinks in the simulations. Instead, it learns the ionization patterns present in the 21-cm signal distribution. Consequently, the framework successfully predicts the accurate volume of ionized regions in simulations generated by \texttt{C$^2$Ray}, a numerical simulation code that conserves photons (\S\ref{sec:ionbubble}). In future work, we plan to retrain the network on models from photon-conserving frameworks, such as \texttt{Beorn} \citep{Schaeffer2023beorn} and \texttt{pyC$^2$Ray} \citep{hirling2023pyc}.

In this paper, \texttt{SegU-Net} was trained on one NVIDIA® Tesla® P100 with 16GB for a total computational cost of approximately 12 GPU hours. When comparing the pre-processing method, we also consider the computational time required to compute the foreground mitigation/avoidance method. In our setup, one lightcone sub-volume of frequency depth $20\,\rm MHz$ with $200$ redshift bins takes about $7\,\rm s$ CPU time to compute with PCA and $2\,\rm s$ with Polynomial fitting. Wedge remover provides faster pre-processing with $230\,\rm ms$ but inefficient foreground mitigation. On the other hand, GPR provides slow but reliable mitigation with a computing time of $\sim 1.2$ CPU hours.

The Grad-CAM importance score analysis conducted in \S\ref{app:blackbox} shows that the network decoder convolutional layer starts by identifying and grouping the region with the strongest positive emission. In the bottleneck of the U-Net model, the low-dimensional latent space then uses the encoded information to identify the threshold that defines the boundary of the neutral regions. The decoder layers use the compressed information and the U-Net skip connection with the encoder layer to define the location of the borders. Finally, the last convolutional layer further refines the decoder output. However, the analysis showed that the network struggled to correctly identify the residual foreground when this signal is similar to the 21-cm intensity. This explains why the final predictions include a positive detection of 21-cm signal regions and a false negative due to the noise or foreground residuals.

In our case, \texttt{SegU-Net} is a deterministic deep learning model. Recently, a series of works have imported probabilistic models in radio astronomy and astrophysics \citep[e.g.]{Friedman2022higlow, Wang2023conditional, Sortino2023radiff}. This approach inherently handles noise and variability in the data compared to the deterministic case. At the same time, they can learn the underlying probability distribution of the data, which can help for a better interpretation. On the other hand, deterministic models like U-Net often have the advantage of being computationally efficient and easier to train. In future work with \texttt{SegU-Net}, we consider converting the model to be probabilistic.

Our analysis shows that using image data from SKA-Low, \texttt{SegU-Net v2} accurately determines the ionization fraction at different stages of reionization. Additionally, we have identified how the ionized regions detected by \texttt{SegU-Net v2} can be used as markers for locating the galaxies responsible for driving the reionization process. These findings demonstrate the potential of our framework for synergy studies with other telescopes, such as the JWST, Euclid and Nancy Grace Roman Space Telescope.

\section*{Acknowledgements}
The authors would like to thank Bharat Kumar Geholt for his useful discussions and comments. MB acknowledges the financial support from the Swiss National Science Foundation (SNSF) under the Sinergia Astrosignals grant (CRSII5\_193826). We acknowledge access to Piz Daint at the Swiss National Supercomputing Centre, Switzerland, under the SKA's share with the project ID sk09. This work has been done in partnership with the SKACH consortium through funding by SERI.  Nordita is supported in part by NordForsk.

The deep learning implementation was possible thanks to the application programming interface of \texttt{Tensorflow} \citep{Tensorflow2015} and \texttt{Keras} \citep{Chollet2017}. The algorithms and image processing tools operated on our data were performed with the help of \texttt{NumPy} \citep{numpy2020}, \texttt{SciPy} \citep{scipy2020}, \texttt{scikit-learn} \citep{scikitlearn2011} and \texttt{scikit-image} \citep{scikit2014} packages. All figures were created with \texttt{mathplotlib} \citep{Hunter2007}.

\section*{Data Availability} 
The data underlying this article is available upon request and can also be re-generated from scratch using the publicly available \texttt{21cmFAST} \citep{Mesinger2011}, \texttt{CUBEP$^3$M} \citep{Harnois-Deraps2013High-performanceCUBEP3M}, \ctworay \citep{Mellema2006CRadiation} and \texttt{Tools21cm} \citep{Giri2020t2c} code. The \texttt{SegU-Net} code and its trained network weights are available on the author's \texttt{GitHub} page: \url{https://github.com/micbia/SegU-Net}.
\bibliographystyle{mnras}
\bibliography{segunet1_mb, segunet1_sg, segunet2}
\appendix

\section{Hyper-parameter exploration}\label{app:hyperpar}
\begin{table*}
    \centering
    \caption{\texttt{SegU-Net} hyper-parameter optimization analysis for the best-performing setups of seven parameters with optimisation on the validation loss.}
    \begin{tabular}{|l|c|c|c|c|c|c|c|c|c|c|}\hline
       Ranking & Activation & Channels latent space & Depth & Dropout & Final activation & Kernel & Pooling type & $r_\phi$ $[\%]$ & Validation loss [$\times10^{-2}$] \\ \hline
        1 & \texttt{LeakyReLu} & 256 & 3 & 0.42 & $\sigma(x)$ & 6  & max     & 89.08 & 6.59 \\ \hline
        2 & \texttt{LeakyReLu} & 128 & 4 & 0.00 & $\sigma(x)$ & 5  & max     & 89.02 & 6.62 \\ \hline
        3 & \texttt{Elu}       & 128 & 3 & 0.34 & $x$         & 11 & average & 87.76 & 7.27 \\ \hline
        4 & \texttt{LeakyReLu} & 128 & 4 & 0.50 & $\sigma(x)$ & 5  & max     & 88.72 & 6.85 \\ \hline
        \textbf{5} & \textbf{\texttt{ReLu}}      & \textbf{256} & \textbf{4} & \textbf{0.05} & \textbf{$x$}         & \textbf{7}  & \textbf{average} & \textbf{88.50} & \textbf{7.48} \\ \hline
        6 & \texttt{ReLu}      & 256 & 5 & 0.14 & $x$         & 7  & max     & 86.53 & 9.15 \\ \hline
    \end{tabular}\label{tab:hyperparams}
\end{table*}
As we mentioned in \S\ref{chap:unetfor21cmimages}, we perform an optimisation analysis of the \texttt{SegU-Net} hyper-parameters. We are aware of tools that automatize the exploratory analysis of the network hyper-parameter space, such as \texttt{Optuna} \citep{optuna_2019}. However, constrained by time and computational resources, we manually searched the best-performing parameters through a trial-and-error approach. First, we selected a few combinations of the network parameters and performed a short training of no more than five to ten epochs. Based on the result obtained in this short training, we selected six of the best-performing results with the lowest validation loss and performed a full training to identify the ideal hyperparameters setup. In future work, we intend to undertake a more comprehensive study.

We list the six best-performing setups we tested in \autoref{tab:hyperparams}. We include an analysis of seven model parameters, from left to right: the activation function of the convolutional layers, the number of channels for the bottom layer, the number of pooling operations, the dropout rate, the final activation function before the binary operation, the size of the kernel filters and the type of pooling operation. As a loss function, we employed the balanced cross-entropy (BCE) function \citep{Salehi2017} and the Adaptive Moment Estimator (Adam) \citep{Kingma2014} as the stochastic gradient descent algorithm to minimise the loss. We employed the in bold text for the results presented in this paper. Although we monitored the validation loss to select the best setup, we noticed that the first and second models gave the worst prediction for images at the edges of the redshift range $z\sim7$ and $11$. The fifth model provided the most balanced result, with an overall $r_\phi\approx0.7$ score, as shown in \autoref{fig:mcc} central panel. Moreover, in contrast to the findings by \citet{Li2018understanding}, we observe that setting the dropout rate to zero enhances accuracy only for the third-ranked setup. Meanwhile, other configurations exhibit improved performance when both batch normalization and dropout are included. 

\section{Inside the Black Box}\label{app:blackbox}
\begin{figure*}
	\includegraphics[width=\textwidth]{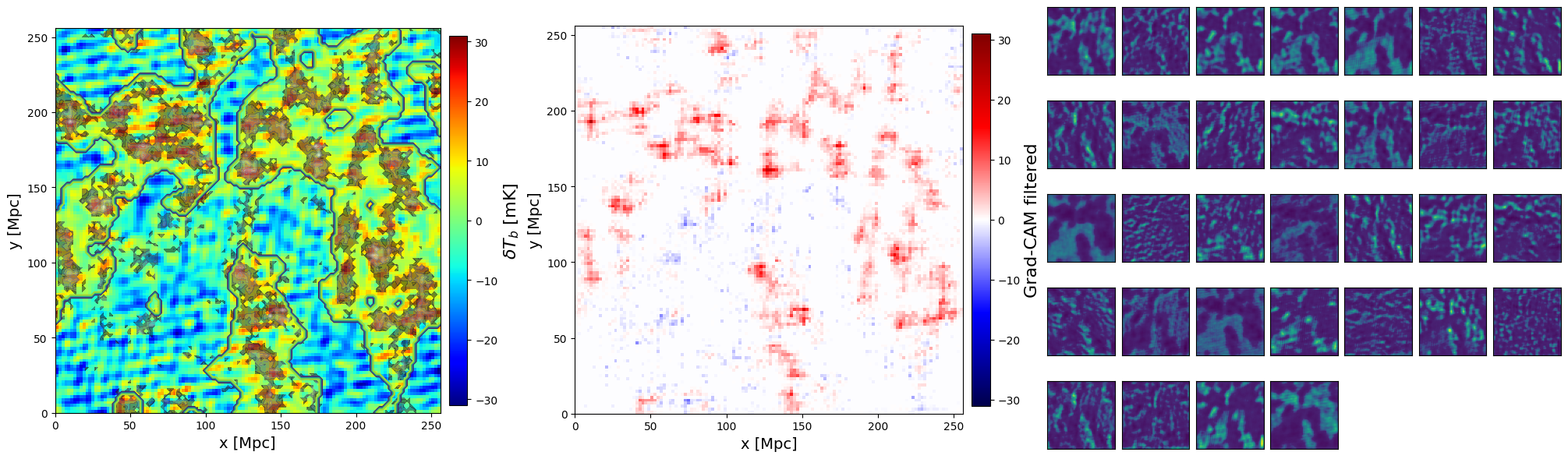}
    \includegraphics[width=\textwidth]{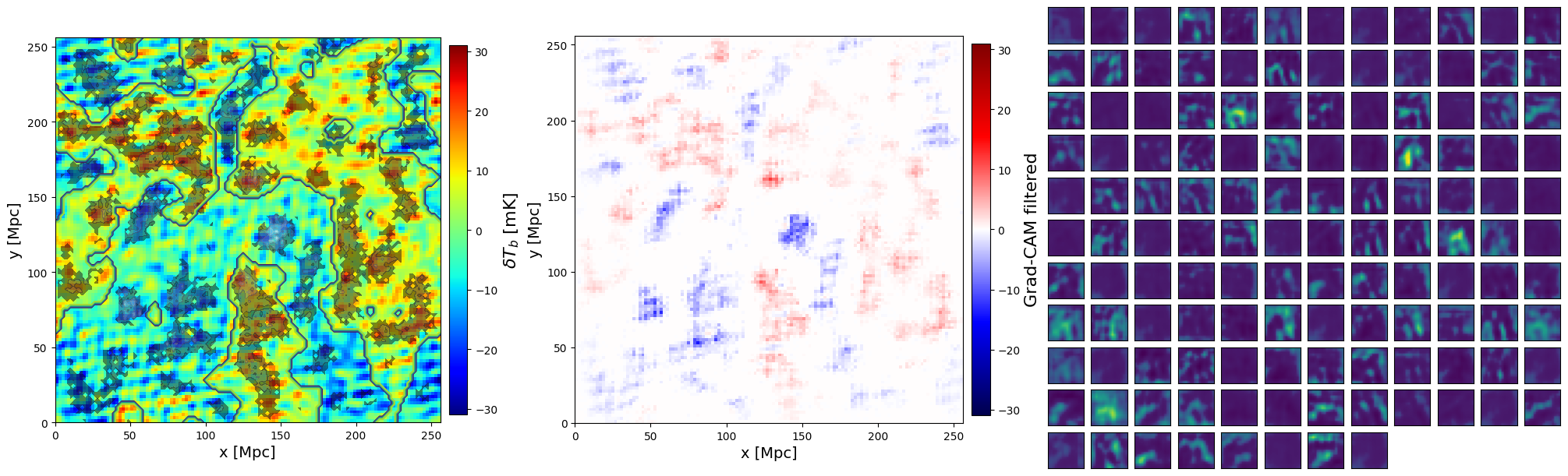}
    \includegraphics[width=\textwidth]{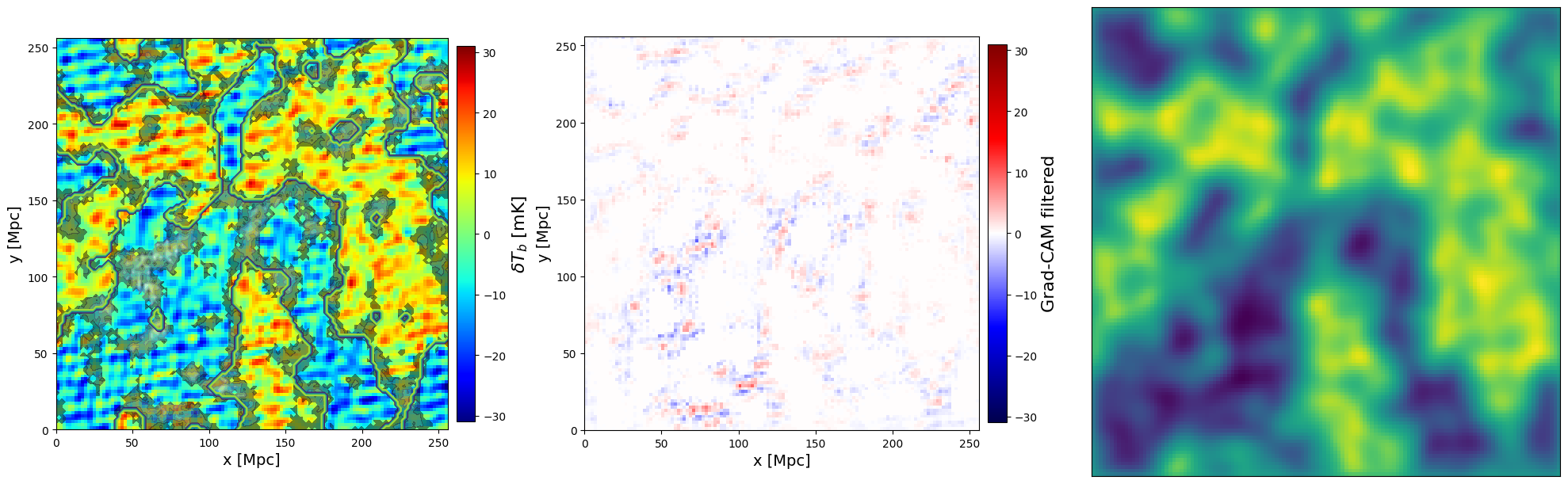}
	\caption{Region of interest detected by Grad-CAM for three hidden layers. \textit{Left panels}: Input image with shadow areas that indicate the region of attention detected by the Grad-CAM method. Black solid contours indicate the ground truth for comparison. \textit{Central panel}: The filtered Grad-CAM image element-wise multiplication between the input and the $M_c$ filter. \textit{Right panel}: A visualisation of the hidden layer output. The number of sub-panels indicates the channel size of the hidden layer.}
	\label{fig:inside}
\end{figure*}
The trained model we presented in this paper is able to recover the ionized field from noisy images with residual foreground contamination. This is an indication that the network learns to identify the regions of interest from important hidden features that maximize the recovery. However, the machine learning model's complexity, high dimensionality and non-linearity make them difficult to interpret and regulate, so these applications are often referred to as a \textit{black box}. Here, we present a first attempt to open and look inside \texttt{SegU-Net} black box. A standard tool to visualize and understand the decisions made by a general convolutional neural network is the Gradient-weighted Class Activation Mapping (Grad-CAM) technique \citep{Selvaraju2019gradcam}. This method applied in segmentation and object classification highlights the features of an input image employed in predicting a particular class. Grad-CAM achieves this by computing the gradients of the predicted class $y_c \equiv x^B_{\rm HI}$ score with respect to the feature maps $F_k$ of all the $k>0$ convolutional layers up to the layer in the network under study. A weighted combination of these feature maps gives the \textit{importance score} $M^{(i,j)}_c\in[0,1]$ which indicates the importance of the feature in the image at location $(i,j)$ for the class $c$th. This score is given as
\begin{equation}
    M_c^{(i,j)} = \frac{1}{Z} \sum_{k} w_k \frac{\partial y_c}{\partial F_k^{(i,j)}} \ ,
\end{equation}
where $Z = \sum_{k} w_k$ is the normalization factor, corresponding to the sum of the $k$th weights, while $w_k$ is the weight corresponding to the feature map $F_k$. In our case, we focus on the neutral regions, categorized with a value of $c=1$ in our maps. 
Values close to one indicate high importance, while in the opposite case, it indicates irrelevance.

In \autoref{fig:inside}, we show the result for three hidden layers. The first column shows the input image and the $M_c$ score represented by dark shadows, indicating the location in the image employed in the classification of the neutral regions. Solid line contours indicate the ground truth. The central column shows the Grad-CAM filtered region obtained by element-wise multiplication of the input image with the importance score. This plot shows us what features the network emphasizes in the image for identifying the neutral region. The right column visualises the hidden layer output, with the number of sub-panels corresponding to the number of channels. 
From top to bottom, we have the output of the convolution block at the second level of the encoder after two convolutional layers and a pooling operation. The hidden state has angular and channel dimensions $(64,\,64,\,32)$. We can see that in the encoder, the network focuses on the regions with the highest intensity, which, thanks to the pre-process presented in \S\ref{sec:frg_mitigation}, are mostly located within the neutral region. The different channels in the hidden layer show a similar conclusion, with the convolutional operation capturing the large-scale region that produces 21-cm signals. In the second row, the bottom of the U-Net, known as the low-dimensional latent space, with dimension $(16,16,128)$, gives a compressed representation of the input image, and it appears to focus on location in the image with the highest and lowest values. Our interpretation is that the network focuses on these extreme values to quantify the "threshold" value that sets the boundary between neutral and ionized regions. This interpretation is also supported by the $128$ hidden layer plots in the right panel, as the compressed data shows different constant values across the channels. In the last row, we show the importance score from the final convolution before the binarization of the output. Here, it appears that the network uses the threshold value defined in the bottom layer of the U-Net to locate the delimitation that defines the neutral regions, as the shadow is located along the contour of the ground truth (black solid line). We notice that some locations in the image with substantial foreground residuals are wrongly included. The hidden layer plot shows that the network struggles to remove the foreground residual completely.


\label{lastpage}
\end{document}